# *LOCAL MEASUREMENTS IN A PARTICLE LADEN JET GENERATED BY A CONVERGENT NOZZLE*

J.A. García, E. Calvo., J.I. García Palacín, J.L. Santolaya, L. Aísa

Fluid Mechanics Group, Universidad de Zaragoza, C/Maria de Luna, Torres Quevedo building, 50015, Zaragoza, Spain.

## *ABSTRACT*

Two-phase flows, either gas-particle or gas-droplet, have a large industrial interest (combustion, medical applications, atomization…).

Because of the complexity of the interaction between phases and the chaotic nature of the continuous phase turbulence and particle motion, it is necessary to study those flows from a basic point of view. This will help to improve their understanding and modeling by stochastic methods (such as Montecarlo).

The aim of this work is an experimental characterization of a particle laden jet, generated by a nozzle. The particles are glass spheres. Phase Doppler Anemometry (PDA) technique has been applied to measure two velocity components and the particle diameter.

In a turbulent shear flow, the Stokes number describes the particle-flow interaction. For large Stokes numbers ($St >> 1$) the interaction of the particle with the gas is due to the mean flow mainly; for small Stokes numbers ($St << 1$) the particle dynamic is strongly influenced by the turbulence.

Near to the nozzle exit zone, the mass flux has been calculated from the raw data (See L. Aisa 2002). This magnitude is the basis to evaluate particle dispersion parameters. The dispersion results are related to the initial conditions and the interaction between phases. These results have been compared with other authors' data [Prevost 1994].

Other subject discussed is the jet turbulence. In particular, particles can modify the turbulence dynamic. So, measurements of the single phase flow are compared with the air flow (continuous phase in the next) of the two phase jet to asses this changes. For the continuous phase measurements, the differentiation between signals coming from glass spheres or tracers, in order to obtain the correct velocimetry for the continuous phase (gas), is critical and it has been discussed.

## *EXPERIMENTAL SETUP.*

A particle laden air jet is discharged through a tube of 1 m of length. The tube ends in a nozzle of 12mm exit diameter with a contraction ratio of 5.14, the exact geometry of the nozzle is shown in fig.1. The tube length allows the efficient mixing of both phases and this nozzle geometry generates a large slip between the phases.

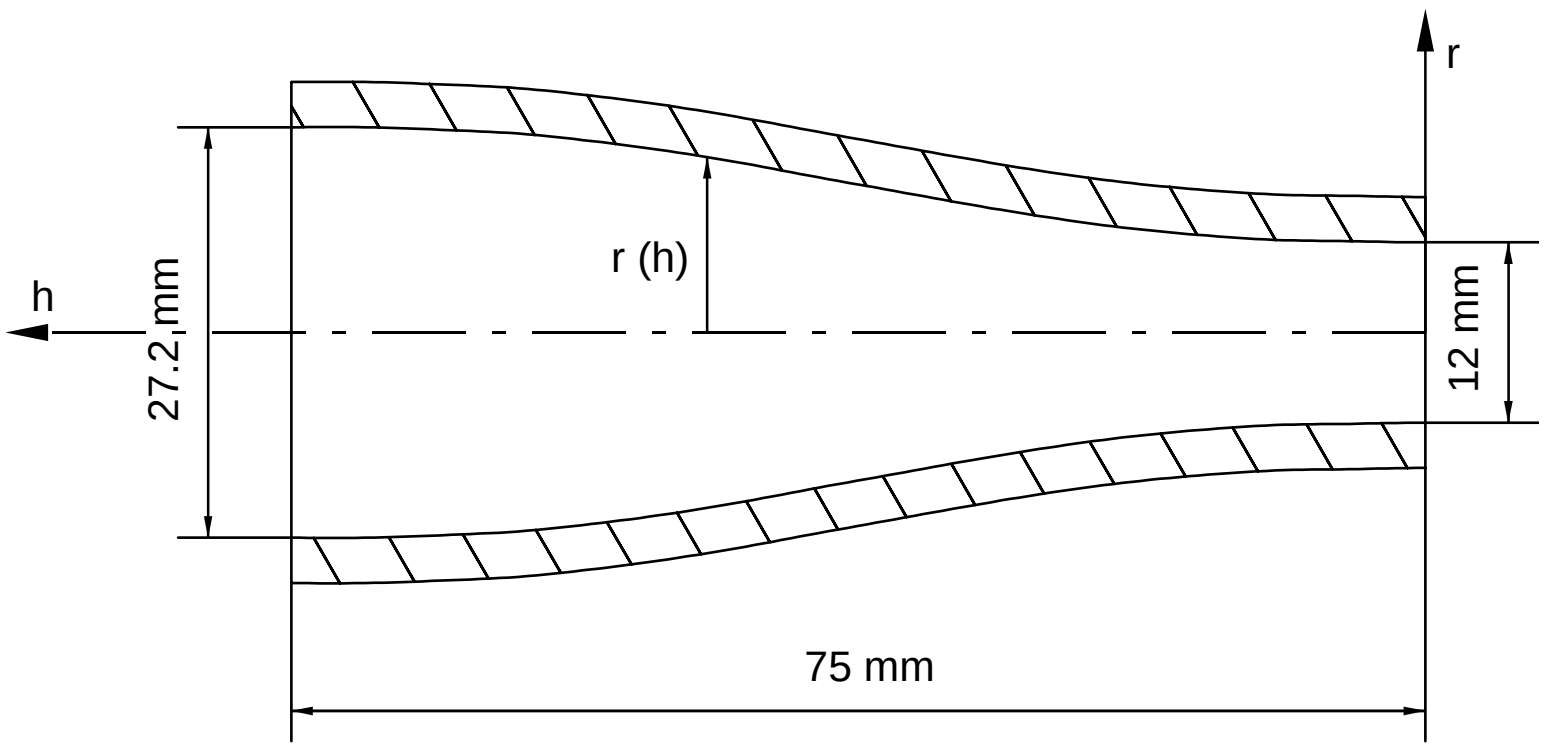

Nozzle profile equation:

$$r(h)=19.22\cdot10^{-9}h^{5}-3.603\cdot10^{-6}h^{4}+180.15\cdot10^{-6}h^{3}+6$$

**Fig. 1. Nozzle geometry.**

A computer drives an electro-pneumatic valve to control the airflow. The air discharges to at atmospheric conditions approximately.

The air is seeded with 0.5 μm diameter alumina tracers for the continuous phase. Its small size allows tracers to follow air movement accurately.

The dispersed phase is made up of glass spheres with a size range between 40 a 110μm, mean diameter of 77.2 μm and a density of 2450Kg/m$^3$. A vibrating device controls the injection of spherical glass particles. In this work the mass ratio (particles/air) is 0.3.

The flow evolves in a square 480 by 480 by 1000 mm transparent chamber. Inside the chamber, there is a low velocity co-flow surrounding the jet.

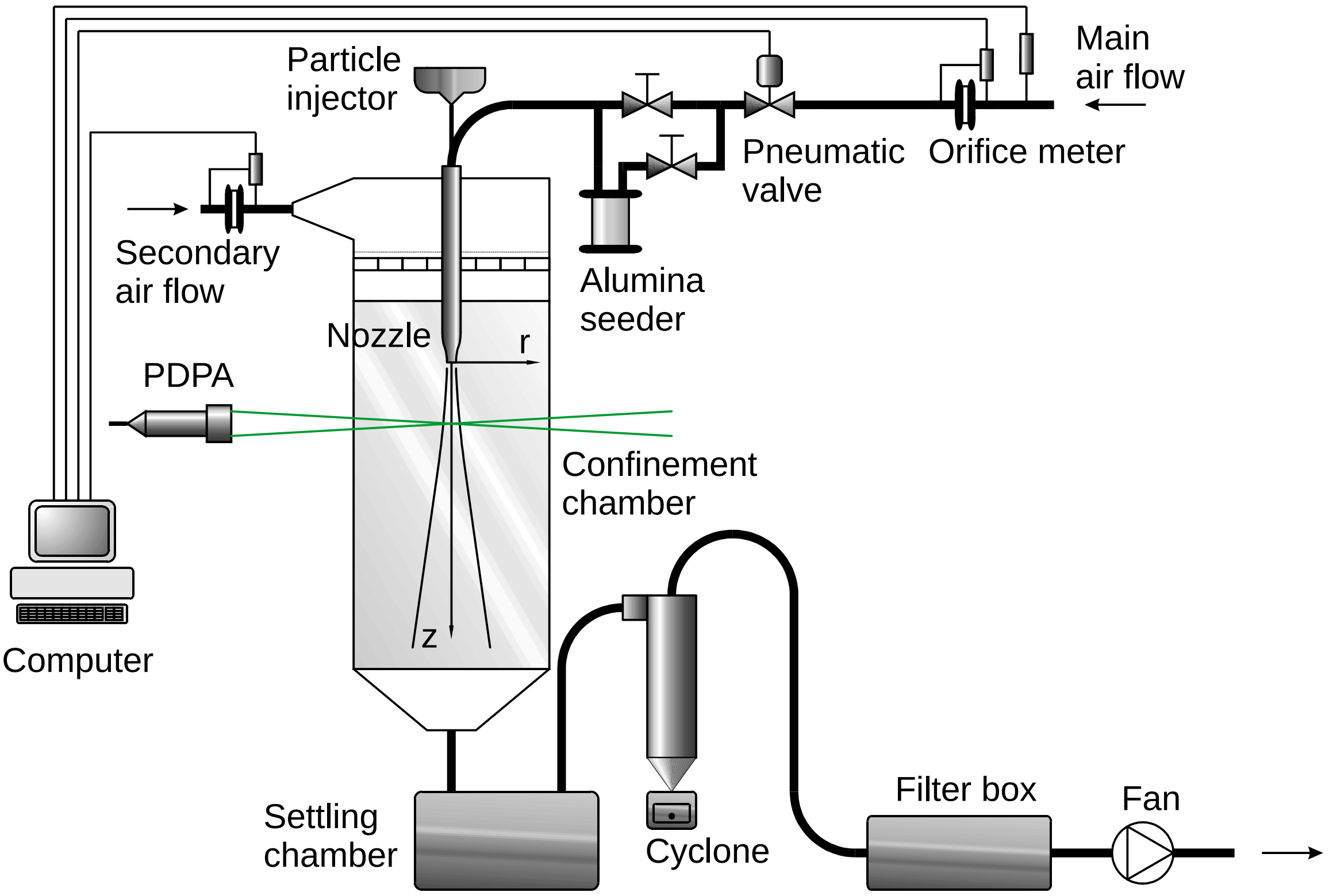

**Fig. 2. Experimental set-up.**

The airflow is adjusted to 1.69 l/s, either for the single phase case or for the continuous phase in the two phase case. The nozzle has been designed to give an exit flat velocity profile for the single phase jet. So, we have a mean exit velocity of 15 m/s, corresponding to a Reynolds number of 12000.

## *MEASUREMENT SYSTEM.*

Phase Doppler Anemometry (PDA) technique has been applied to characterize the flow experimentally. The system uses two pairs of beams (a pair with a wavelength of 514.5nm and the other pair with a wavelength of 488 nm). The probe volume is the region defined by the interference region between both laser beams of each pair.

The system measures two velocity components of each particle which crosses the probe volume and its size if the particle is spherical, as principal parameters. Also, it provides the time when a signal starts, the time that the particle needs to cross the probe volume (named transit time) and the intensity of each signal. With this information is possible to calculate the local mass flux or the concentration of particles (See thesis of J. A. Garcia 2000 and L. Aisa 2002).

Only if a particle is spherical its diameter is accurately measured by PDA. In this case, the glass particles are spherical. However, the tracer particles (alumina) have an irregular shape and its measured diameter is wrong. Fortunately, the diameter distribution of the alumina tracers given by the PDA for the alumina particles is out of the range for the glass particles.

A typical distribution for the alumina particles is shown in the figure 3.

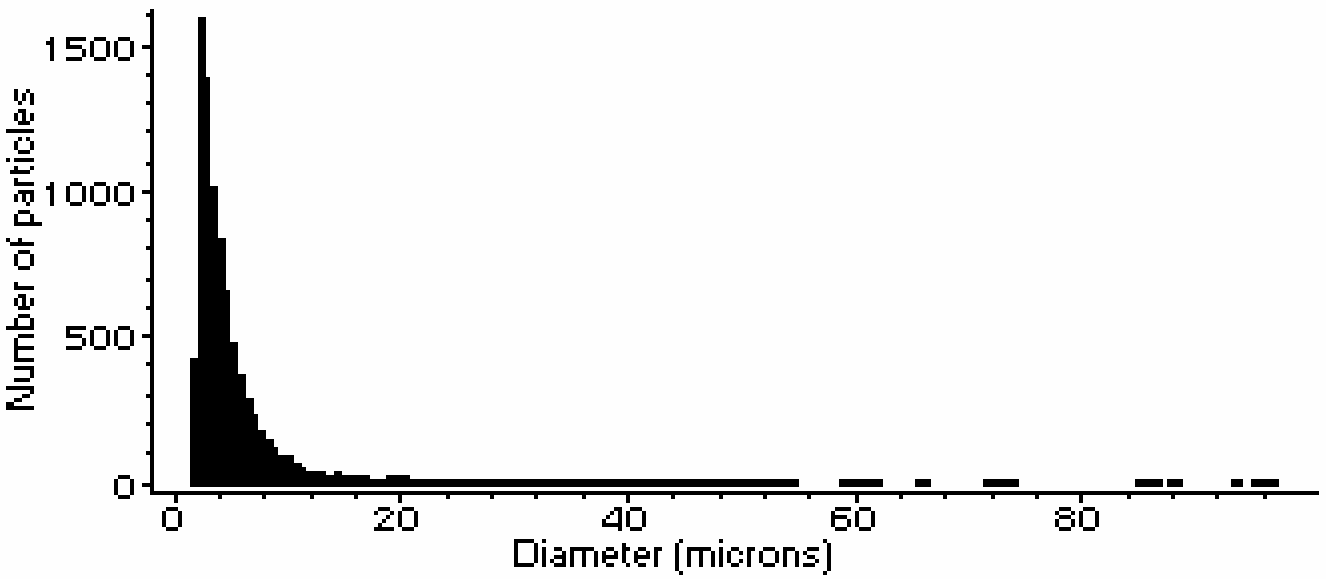

**Fig 3. Measured diameter distribution for the tracer particles.**

For the glass particles, we have distributions like the shows in the figure 4.

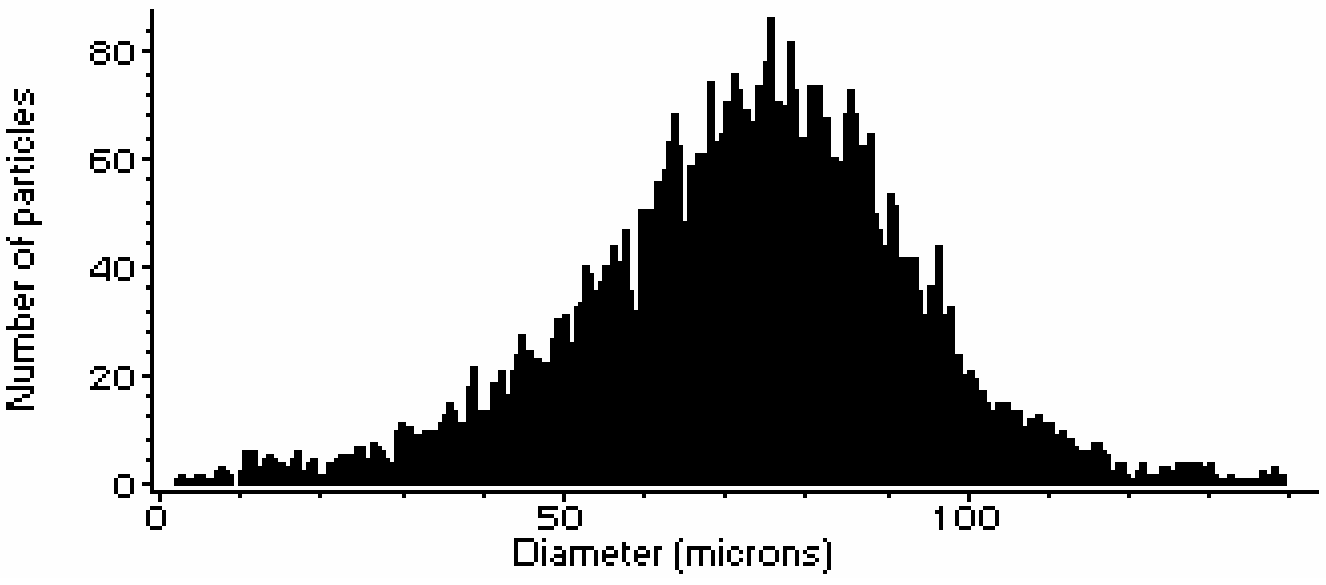

**Fig 4. Measured diameter distribution for the glass particles.**

A distribution with both types of particles appears in the figure 5.

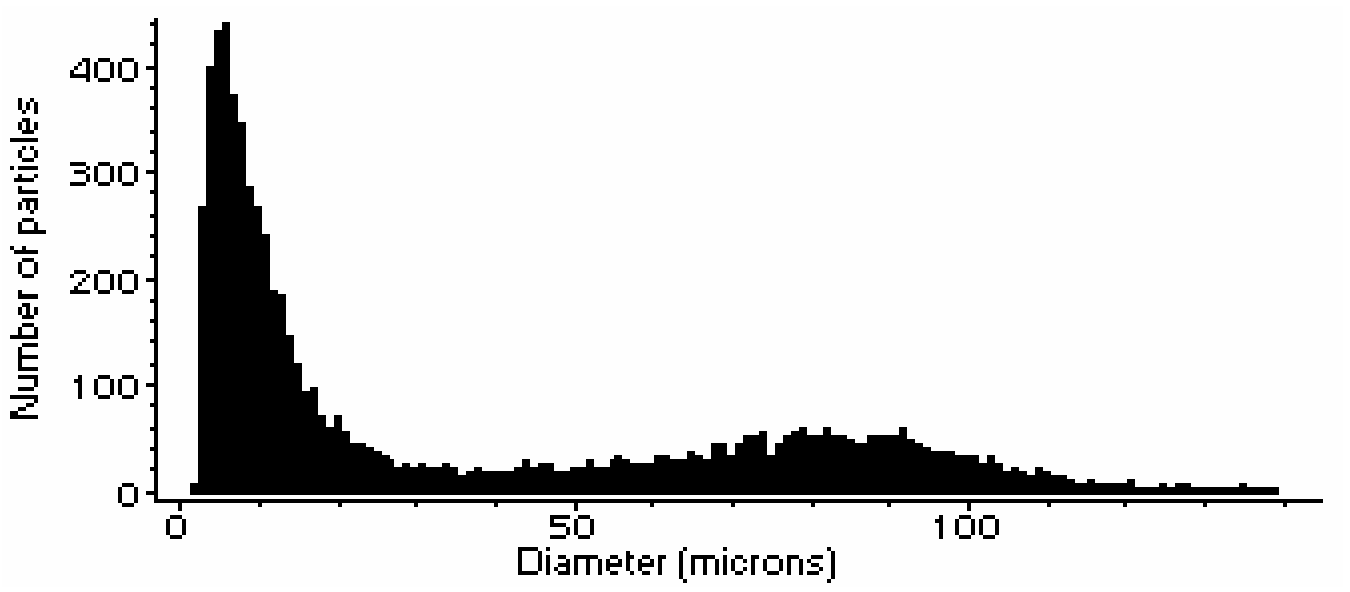

**Fig 5. Measured diameter distribution for both types of particles.**

As the reader can notice, there are some large diameter for alumina particles, and some small diameters for glass particles. In order to obtain the statistics of the glass particles, the experience was made without tracer particles. To get the data for the continuous phase, only the signals with a diameter lower than 10 μm have been considered. Besides, among these signal, those with high intensity have been rejected. So, the statistic of the continuous phase has been built with particles whose diameter is lower than 10 μm and with low signal intensity.

## *INITIAL CONDITIONS AND VISUALIZATION.*

The velocity profiles at the section z/D=0.5 are taken as the initial conditions, because it is impossible the measurement at the exit section with the PDPA system.

The velocity and fluctuation profiles for both phases, and the comparison with the single phase jet, appear in the graph 6:

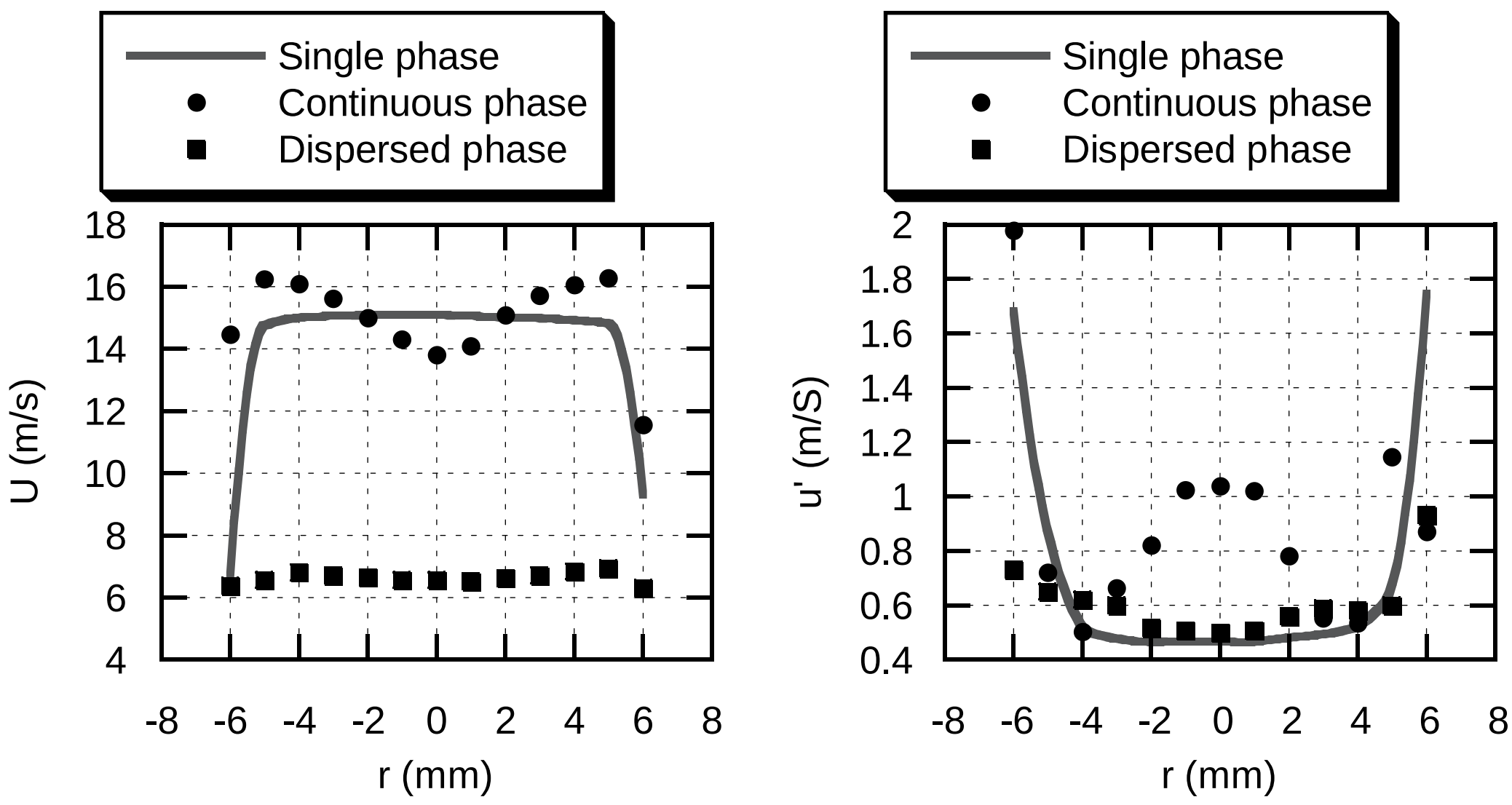

**Fig 6a, b. Mean axial velocity and axial turbulent intensity at the z=0.5D section.**

Comparing the single-phase and continuous phase velocities, the particle presence has a weak influence in the mean velocity of the gas. There is a small velocity lack in the jet axis due to the particle drag. However, the addition of particles gives an important increasing of the fluctuation in the central zone, sometimes times called core jet.

It is interesting to study the profile of particle volume flux to understand the changes due to the particle presence.

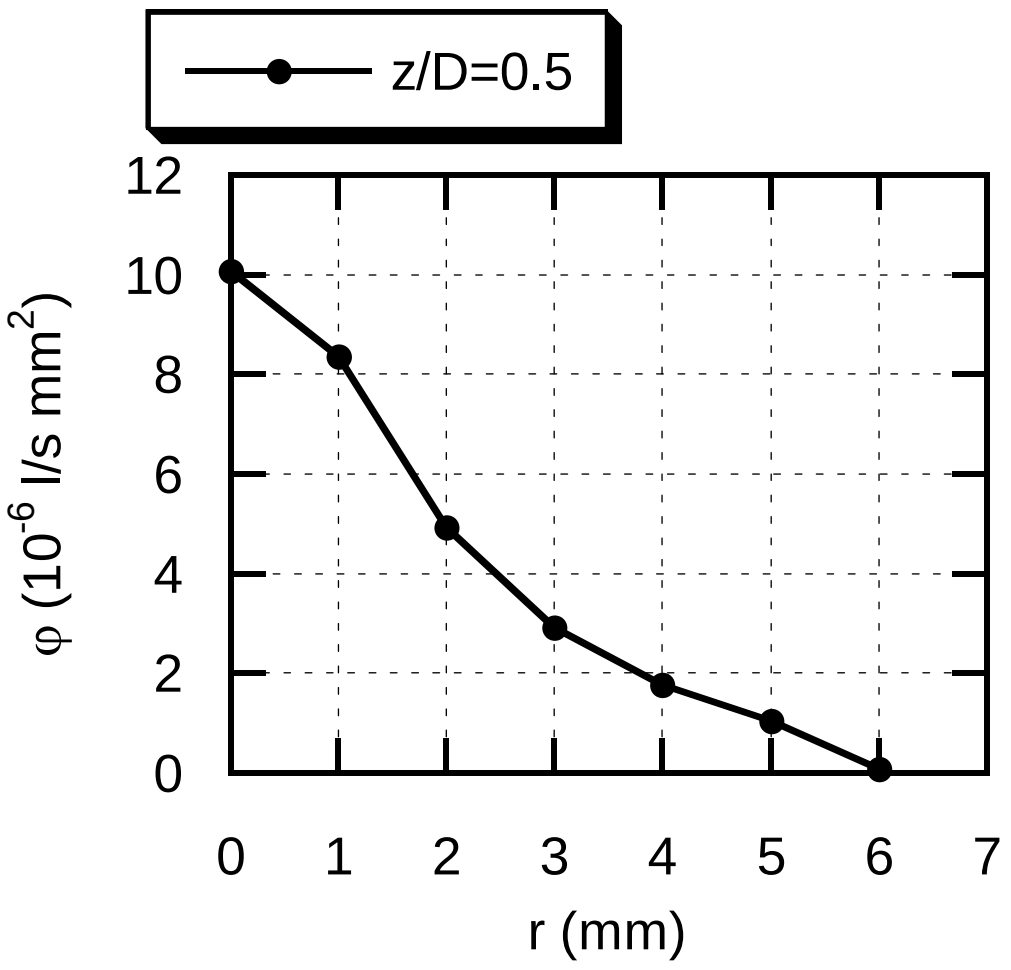

**Fig 7. Mass flux semi-profile at the section z=0.5D.**

The focusing of particles in the central zone is caused by the geometry of the convergent nozzle and the rebounds of the particles on its wall.
The nozzle induces a strong slip between the dispersed and the continuous phase but, also, it is responsible of other effects, mainly:

- The collisions of the particles with the internal nozzle wall leads to the particles focusing in the central zone. Also, these collisions can induce particles rotation.
- The convergent nozzle reduces the turbulence. This is clear for the single phase case but for the continuous phase this effect competes with the fluctuation that is induced by the particles.

Figures 8a and 8b are the jet visualization.

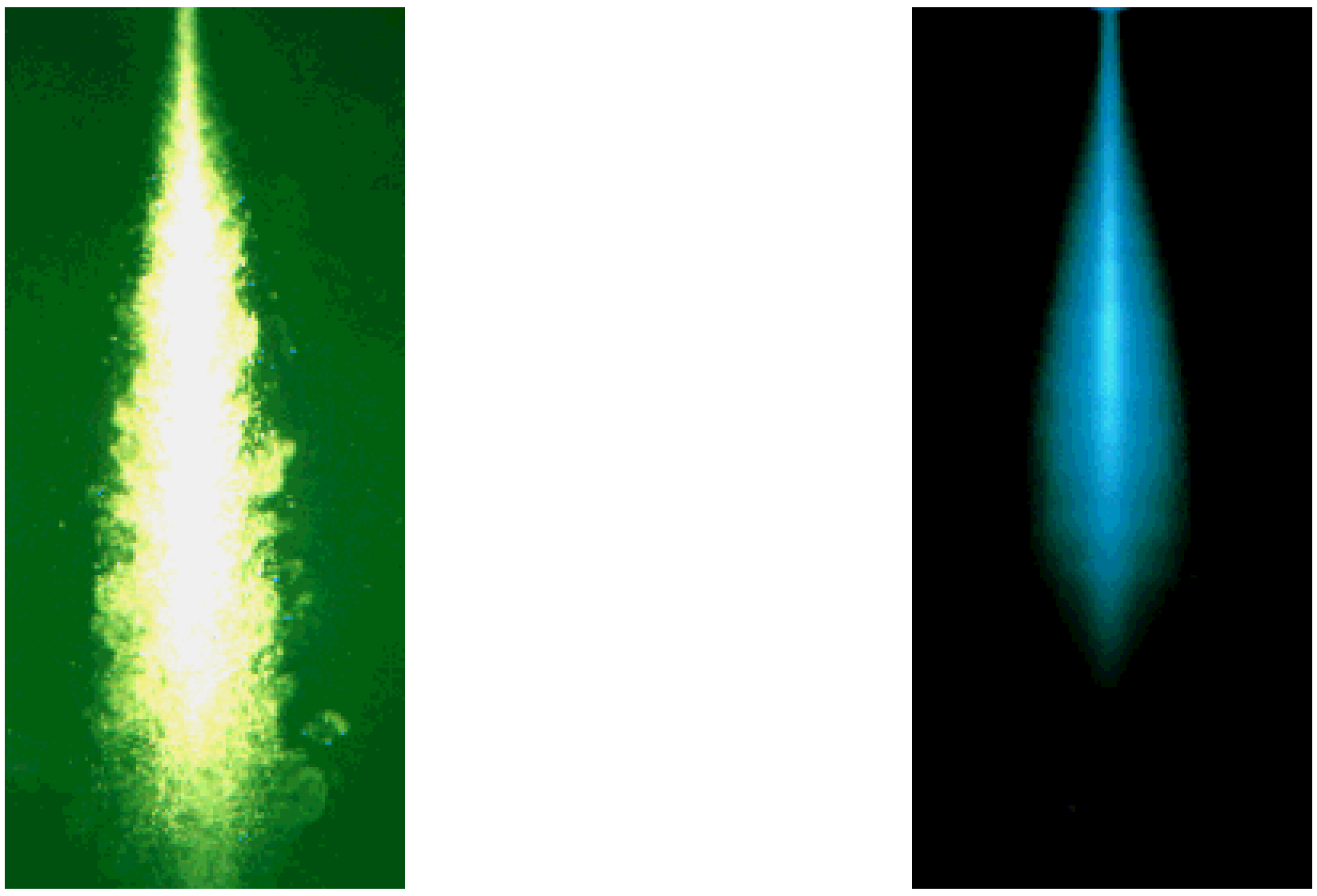

**Fig 8 a, b. Instantaneous and averaged image of the particle laden jet.**

The images have obtained using an Nd-Yag pulsating laser for instantaneous case and a continuous $Ar^+$ laser for the averaged image.

## *EXPERIMENTAL RESULTS.*

Massive data have been obtained for both phases in the two phase case and for the single-phase jet.

Axial and radial velocity components, their fluctuations and their correlation have been acquired along the axis jet and across several transversal sections. This information is enough to compare with previous referenced works about jets generated by straight pipes and for a detailed cinematic description related to the jet characteristic scales.

The studied zone is split in two zones: i) the near to the nozzle zone ($z/D < 10$), where there is a quick development; and ii) the far to the nozzle zone ($z/D \geq 10$), where there is a slower development. Also, it can be seen that in the near to nozzle zone the continuous phase have a larger velocity that the particles and in the far to the nozzle zone the continuous phase is slower than the particles, so the particle drag due to the continuous phase acts in the opposite direction. In the first zone, the study is more detailed, there are information classified by particle diameter and mass flux data is provided.

### <u>1- Axial development.</u>

The next graphs show the development along the axis of the mean axial velocity and its fluctuation.

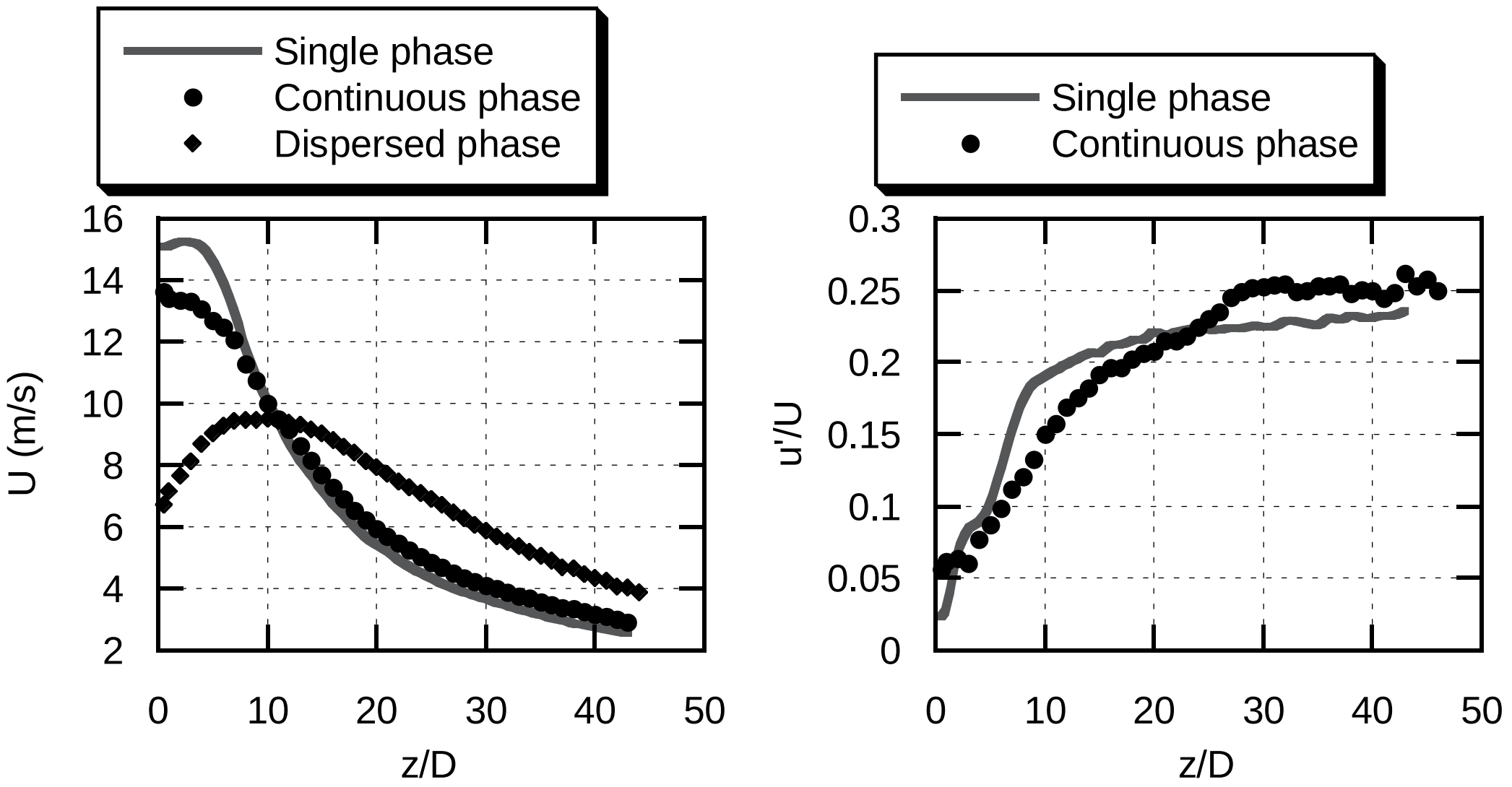

**Fig 9 a, b. Axial development of the mean axial velocity and its fluctuation.**

- Development description.

In the near to nozzle zone the gas phase is modified by the presence of particles. The potential core that appears in the single phase case disappears for the particle laden jet. In the advanced zone (far to the nozzle), the velocity values are different for the continuous phase and the single phase case, but the shape is similar. This effect is discussed later. The particle mean velocity is increased in the first development zone due to the drag that induces the continuous phase.

The fig 8b shows that the starting value of the axial velocity fluctuation is larger for the continuous phase than for the single phase case. As far as z=20D the particle presence delays the fluctuation development; but from z=30D the value of fluctuation is larger in the loaded case and it seems to reach the stabilization.

The effects which govern the velocity evolution for the continuous phase are:

- The exchange of momentum with the dispersed phase.
- The transference of energy that is associated to turbulent structures.
- Local modification of the flow in the particle wake.

The phenomena that govern the velocity field of the dispersed phase are:

- The initial momentum.
- The aerodynamic drag (mainly because of the mean flow in this case).
- The gravity force.
- Magnus effect.
- Collisions between particles.

- Interpretation for the continuous phase.

The strong reduction of the axial velocity of the continuous phase at the axis, compared with the single phase case, is due to the drag of the particles, and to other effects as the turbulent modification.

The velocity attenuation is similar to the one of the single phase case, but with larger values due to the momentum exchange with the particles, faster than the gas in this zone.

The next graph shows the development of the dimensionless inverse of the velocity. This representation is usual to determine the virtual origin.

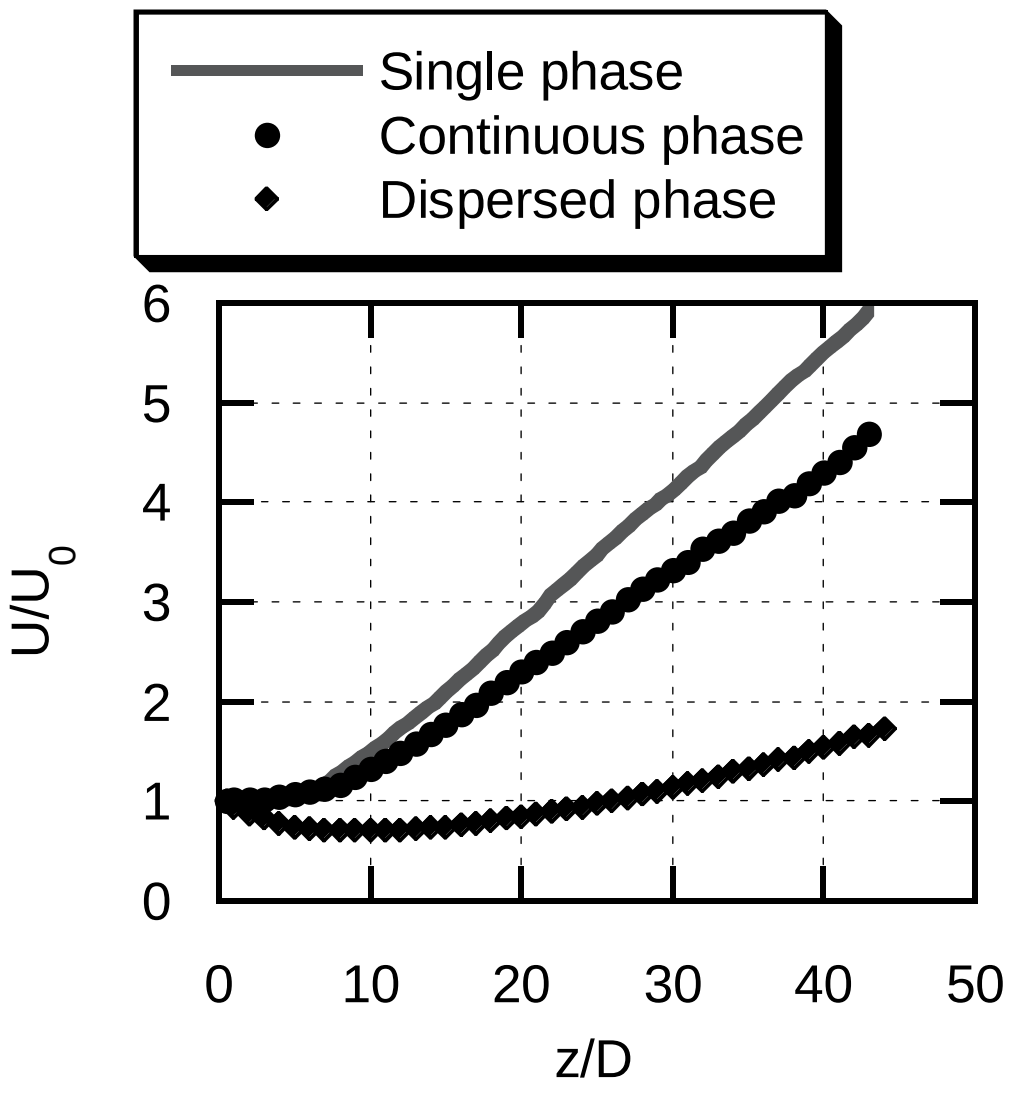

**Fig 10. Inverse axial velocity development.**

The matching to a straight line (for $z/D \geq 20$) gives:

| CASE: | Equation for the straight line |
|---|---|
| CONTINUOUS PHASE | $U_o/U = 0.322 + 0.0997(z/D) \quad R = 0.9994$ |
| SINGLE PHASE | $U_o/U = 0.031 + 0.137(z/D) \quad R = 0.9999$ |

For $z/D \geq 20$, it can be written.

$$U_o/U = K\,(z/D - z_0/D)$$

$z_0$ is called virtual origin. For our case we have:

- Single phase case: $z_0/D = -2.72$
- Continuous phase: $z_0/D = -3.23$

Two conclusions can be obtained about the virtual origin:

1) The virtual origin is slightly farther when there are particles.

2) For this nozzle, the virtual origin is inside the nozzle. In works with tube, the virtual origin is out of the tube (for example: $z_0 / D = 4$, Prevost 1994).

- Interpretation of the particle mean velocity development.

The particle acceleration in the first development zone is due to strong slip between phases. In the advanced zone, the gas reduces the particle velocity by drag and appears the influence of other effects as the turbulence. To evaluate the influence of the turbulence in the particle movement, it is usual to use the dimensionless Stokes number. It is defined as the division between the characteristic time of the particle, $\tau_p$, and the characteristic time of the turbulence, $\tau_g$.

$$S_t = \tau_p / \tau_g$$

There are two asymptotical cases:

- For $S_t << 1$ the particles follow accurately the gas fluctuations. It is the case for the tracers.
- For $S_t >> 1$ the particles are driven by the mean flow only. They are not influenced by the turbulent fluctuations.

Usually, for a jet is written as:

$$St = \frac{\frac{1}{18}\frac{\rho_p D_p^2}{\mu}}{\frac{b_{1/2}}{u'}}$$

Where

$\rho_p$ is the particle density.

$D_p$ is the particle diameter.

$\mu$ is the dynamic viscosity.

$b_{1/2}$ is the radial distance where the axial velocity goes down to the half of the central axial velocity (Usually, it is taken as the characteristic scale of the turbulent macrostructures).

$u'$ is the axial velocity fluctuation.

However, in our case, we have made corrections: i) for the characteristic time of the particles; ii) the characteristic time of the flow is substituted by the interaction time, which means the time that the particle is influenced by a turbulent structure.

The estimation of the Reynolds number for the particles is made to characterize the drag. The calculus is made for the mean diameter.

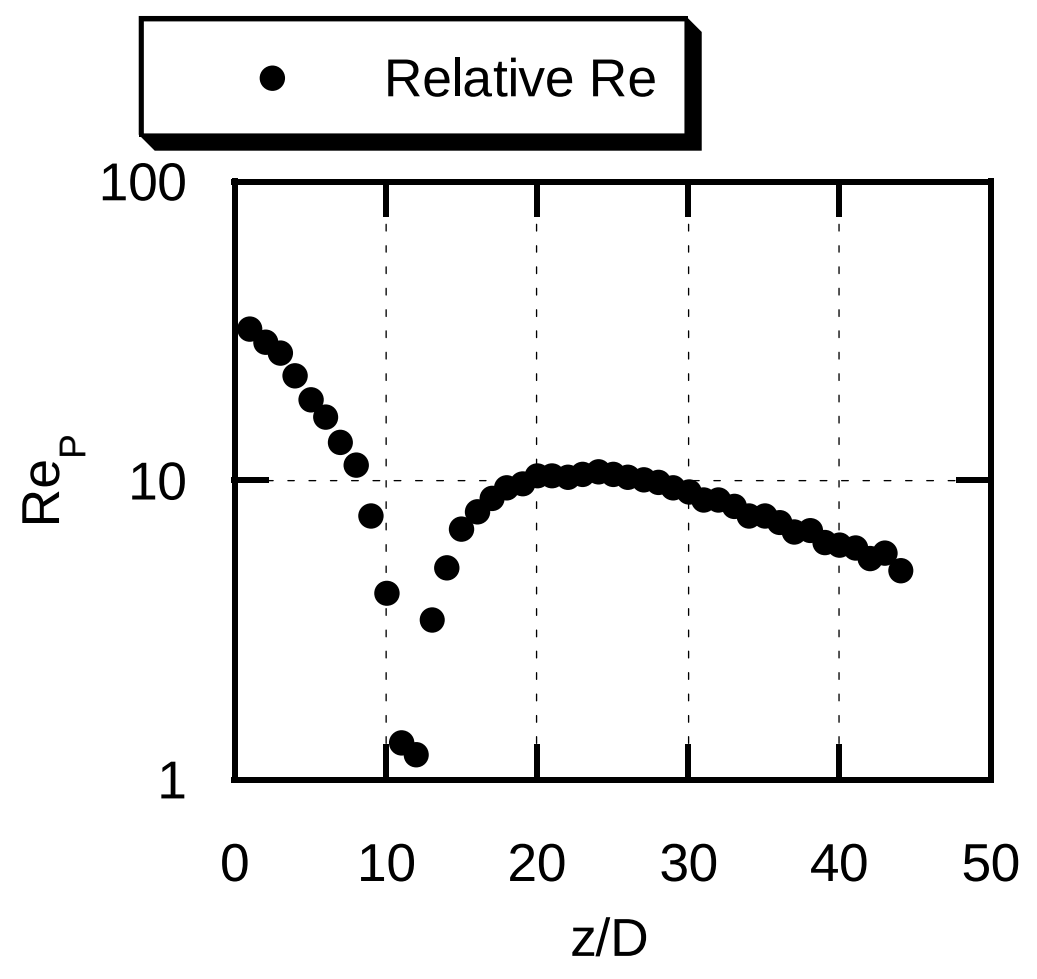

**Fig 11. Particle Reynolds number development.**

In order to have a more representative description of the coupling between the particles and the turbulent structures, we have made some corrections:

- We have evaluated the drag coefficient as $C_D = \frac{24}{\mathrm{Re}_p}\left[1+\frac{1}{6}\mathrm{Re}_p^{2/3}\right]$, following Klyachko (1934).

- For the interaction time the expression proposed by Deutch y Simonin (1991) is used:

$$\tau_{\text{int}} = \tau_g \frac{1}{\left(1+\left(\frac{U_r}{u'}\right)^2\right)^{1/2}}$$

So, we can obtain the corrected Stokes number for the mean diameter, whose development is shown in the figure 12.

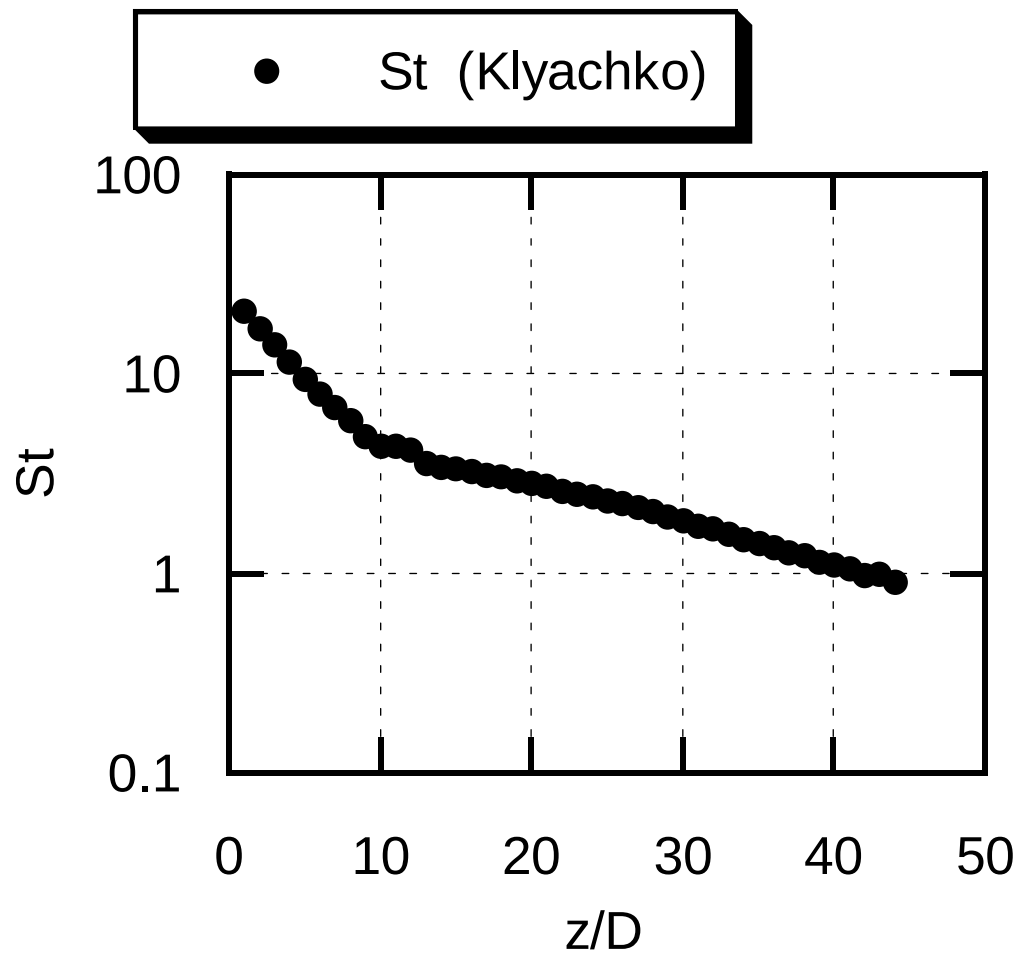

**Fig 12. Stokes number development.**

So, the influence of the turbulent structures is weak, except for the farthest to the nozzle zone ($S_t \approx 1$).

Also, the presence of particles slows down the development of the continuous phase.

2- Transversal sections.

*Near to nozzle zone:*

*A) Continuous phase.*

For the continuous phase, the measurements have been made in the central zone of the jet. We have obtained the next profiles:

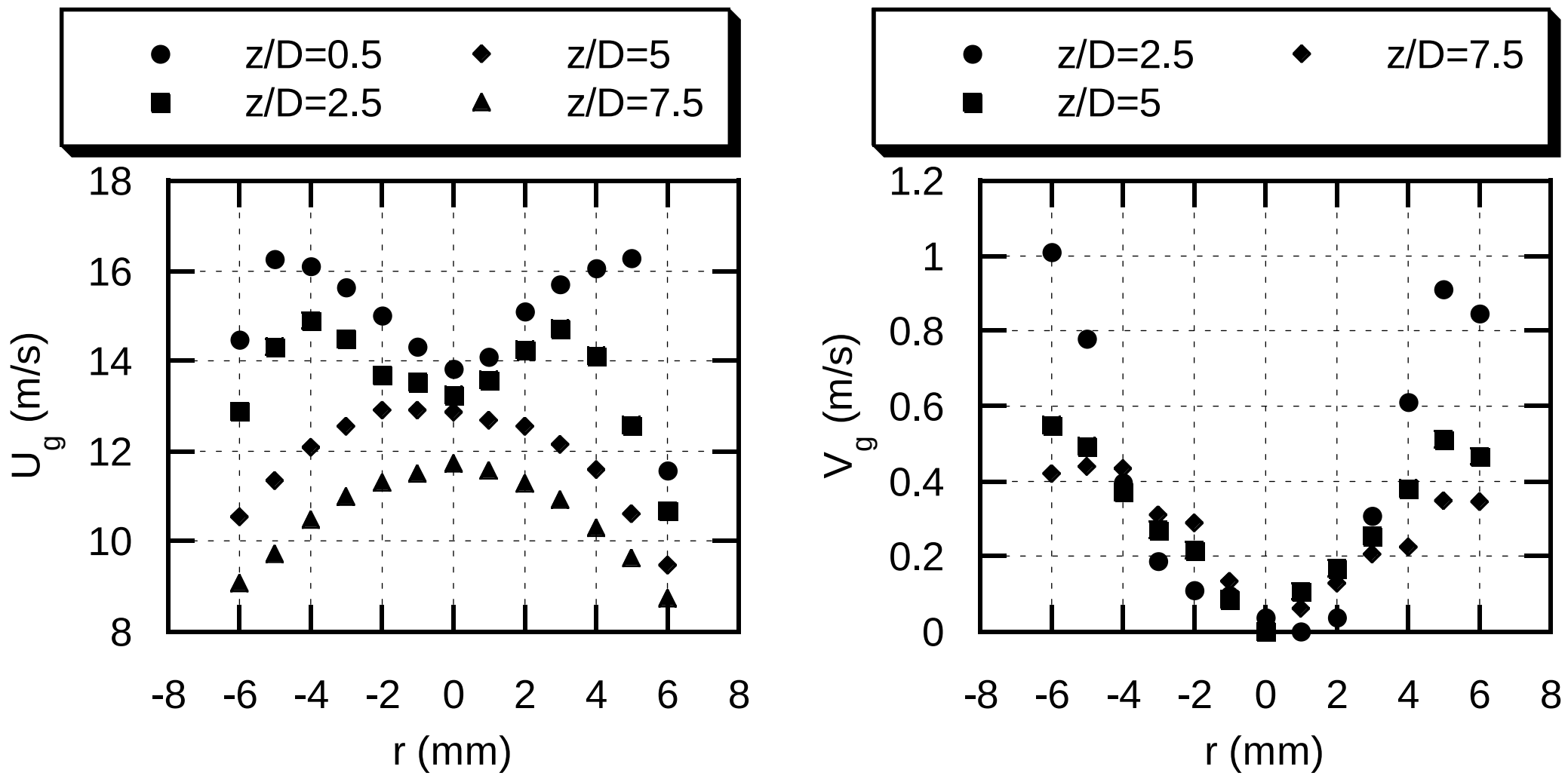

**Fig 13 a, b. Axial and radial velocity profiles.**

If we compare with the single phase case, it can be seen a deficit of velocity in the central zone for the sections at z= 0.5 , 2.5D due to the particle presence and aerodynamic drag. The profiles shows a maximum in the central zone for the next sections, showing a quick flow development to a profile characteristic of a shear flow. The mean radial profiles show the characteristic expansion of a shear flow.

The second order momentum can be observed in the figures 14a ,b.

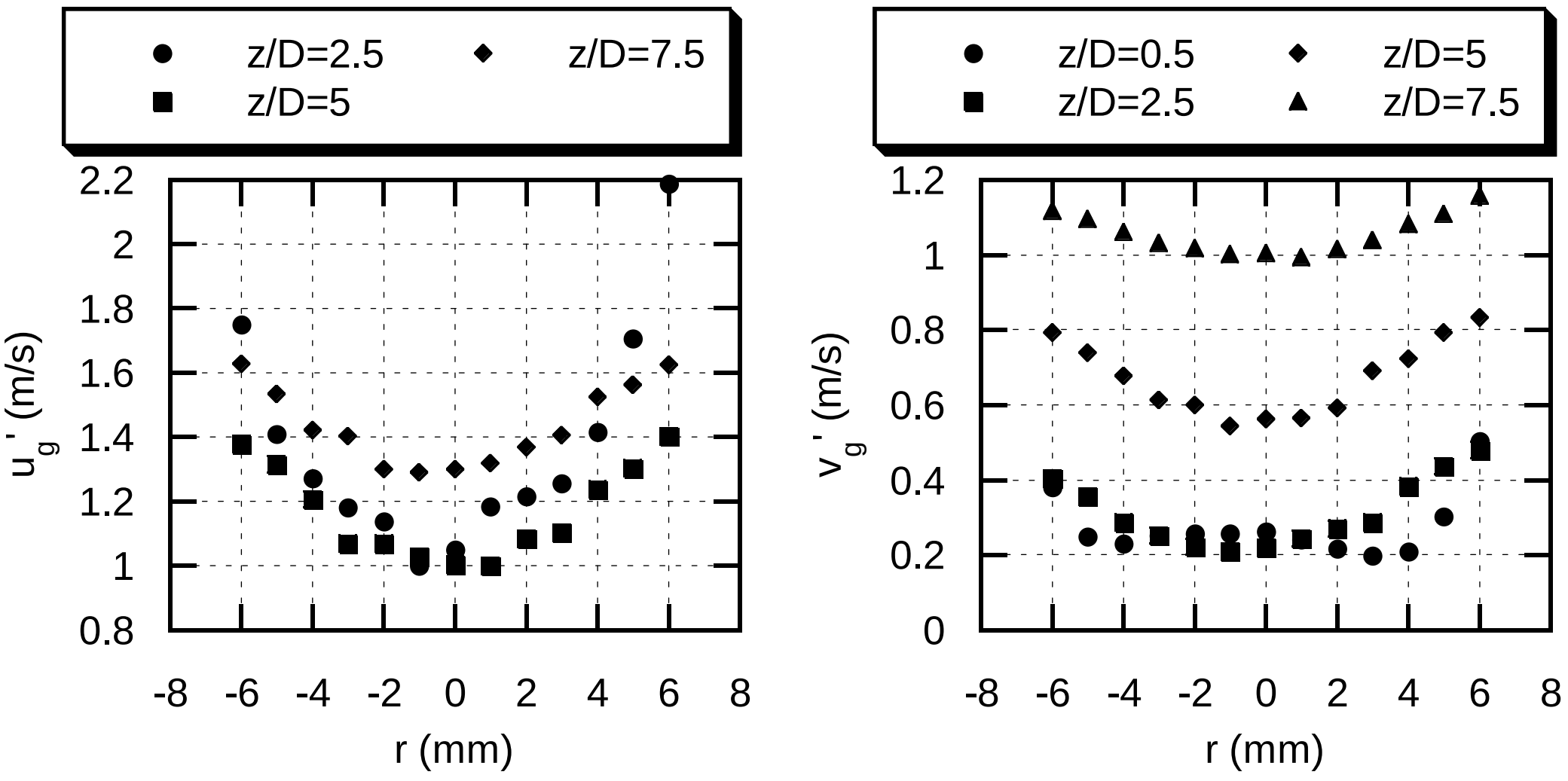

**Fig 14 a, b. Axial and radial velocity fluctuation profiles.**

The central peak in the axial fluctuation in the section z=0.5D disappears in the following sections (See fig. 6a) and the development is slow with a smooth central minimum.

The fluctuation is larger for the axial component, but the development is faster for the radial component.

*B) Dispersed phase.*

The velocity profiles for the particles in the sections $z/D = 0.5, 2.5, 5, 7.5$ are shown in the next graphs.

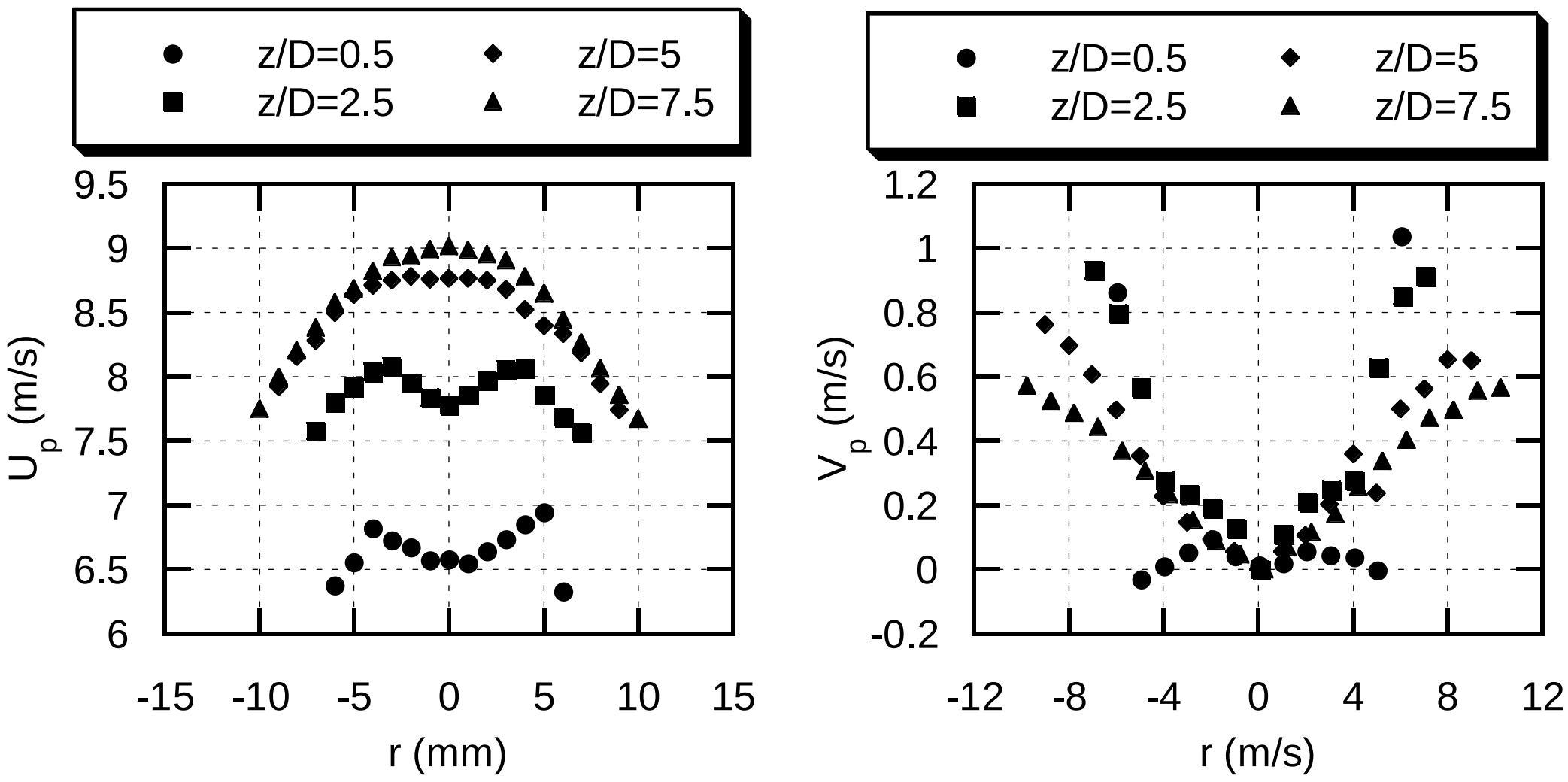

**Fig 15 a, b. Axial and radial particle velocity profiles.**

The particle mean axial velocity has a local minimum in the core jet for the sections z=0.5, 2.5D. Downstream, this profile evolves to a convex shape. There are not large axial or radial velocity gradients in the central sheets. Also, we have seen that the turbulence is very weak near to nozzle and its influence very small ($S_t >> 1$). So, the particles of the same diameter with similar exit direction follow with similar trajectories in the first development zone. Following, this effect will be named "directional classification".

For the radial component, there is a quick development, starting from values closes to 0. The values for the radial component are an order of magnitude smaller than the corresponding to the axial component.

For the fluctuation, we have obtained the profiles in figures 16a and 16b.

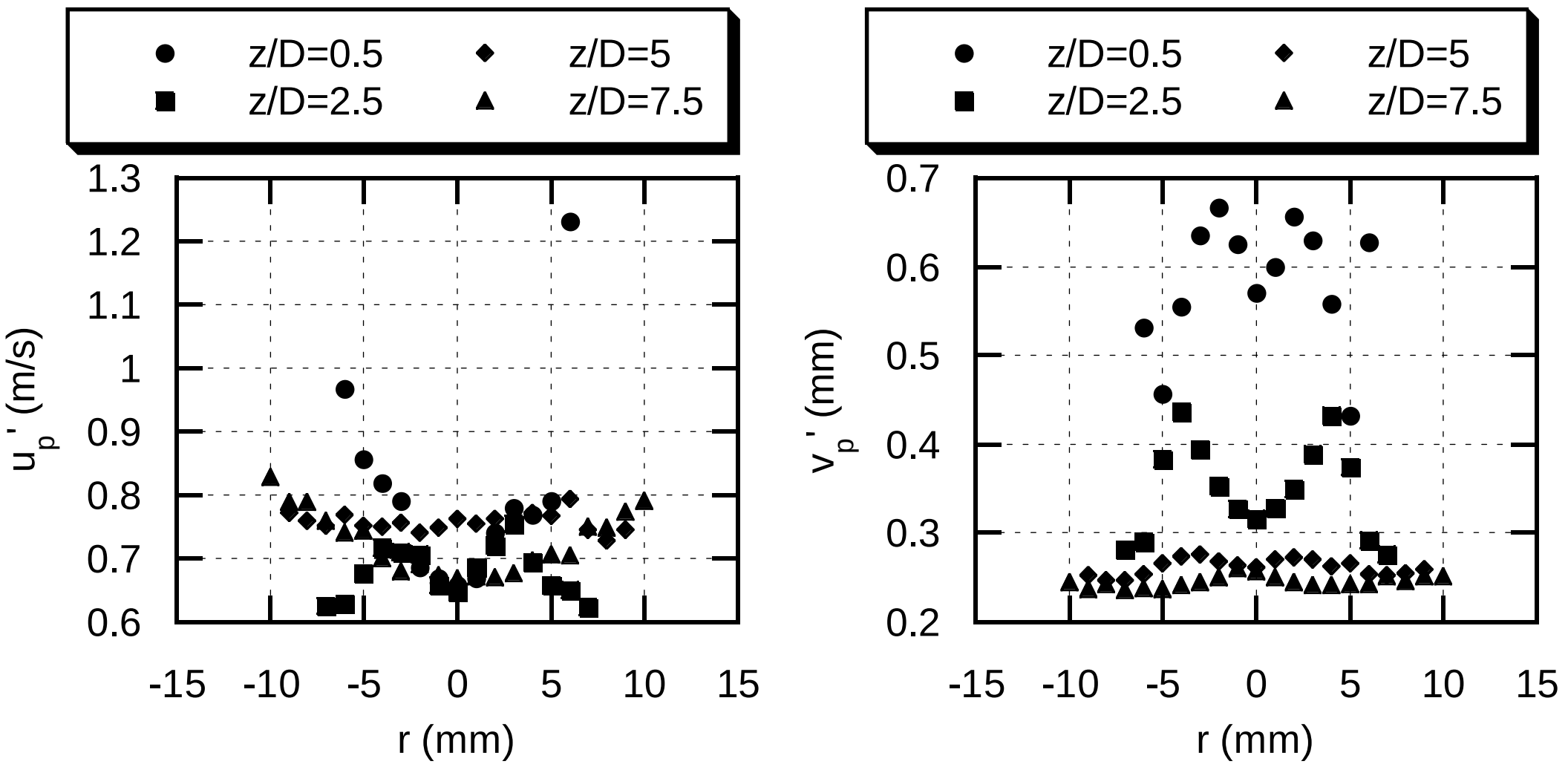

**Fig 16 a, b. Axial and radial particle velocity fluctuation profiles.**

The evolution of the axial fluctuation is very weak. However, for the radial fluctuation we have a strong attenuation and smoothing. This is congruous with the effect of directional classification.

The development of the correlation between the axial and radial component, $\overline{uv}_p$, is shown in the next figure. It is also driven for the directional classification. There are values of $\overline{uv}_p$ negatives in the first sections, which imply a momentum contribution from the external to the internal sheets.

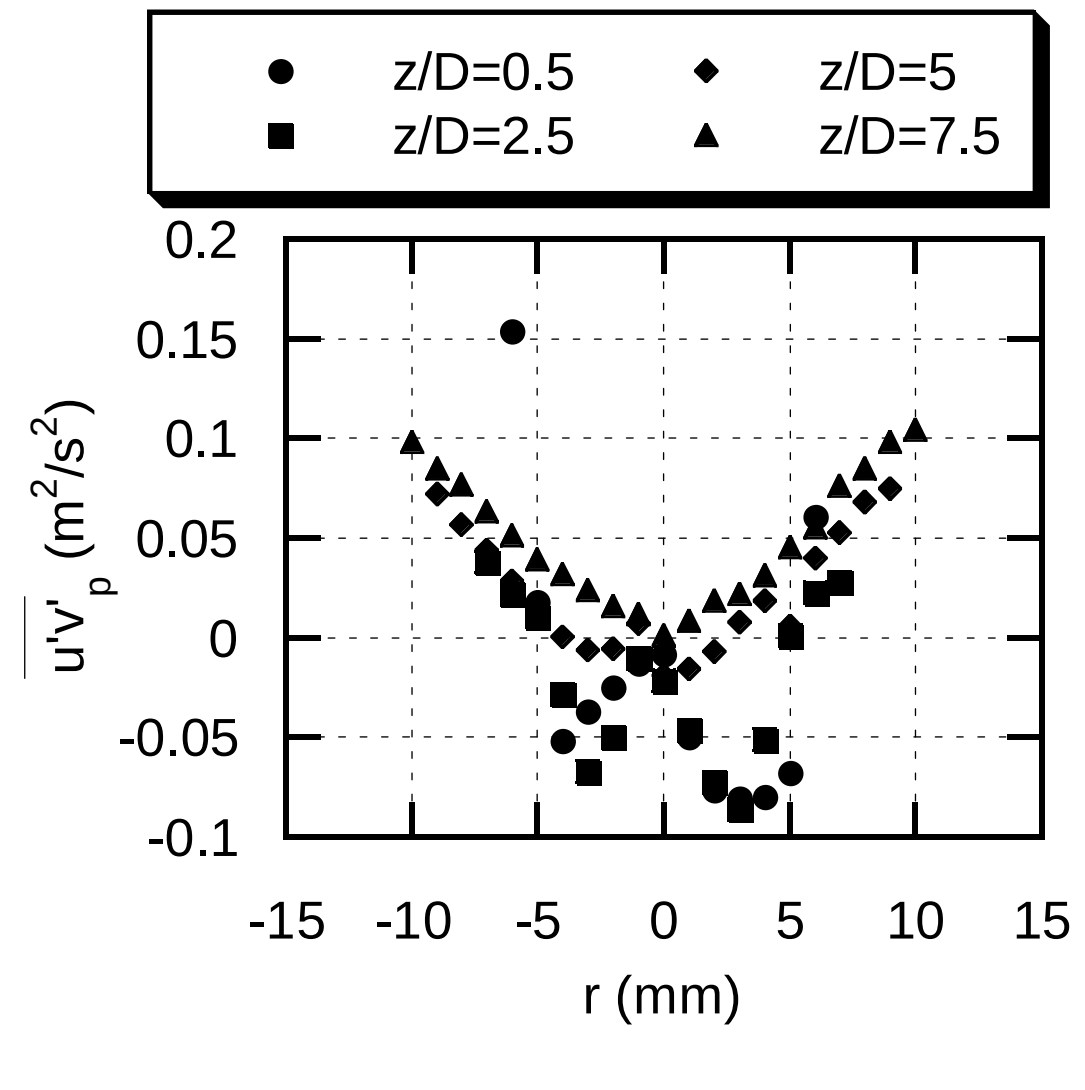

**Fig 17. Axial and radial velocity correlation profiles.**

The comparison with figure 15b allows to think that the bigger particles have a positive radial velocity in the z=2.5D section, but the smaller particles (with the larger axial velocity) have a negative radial component. So, a slight focalization for the z=2.5D section remains still for the smallest particles.

A significant amount of the injected particles collides with the nozzle wall. So, the initial radial component of the particle velocity is very influenced by this effect. To characterize this effect we have obtain the particle direction distribution. Figures 18a, b, c, d and 19a, b, c, d, e, f show the directional histograms at some points of the sections z=2.5 and 7.5D respectively. In these figures, $\psi$ is the angle between the particle direction and the z axis.

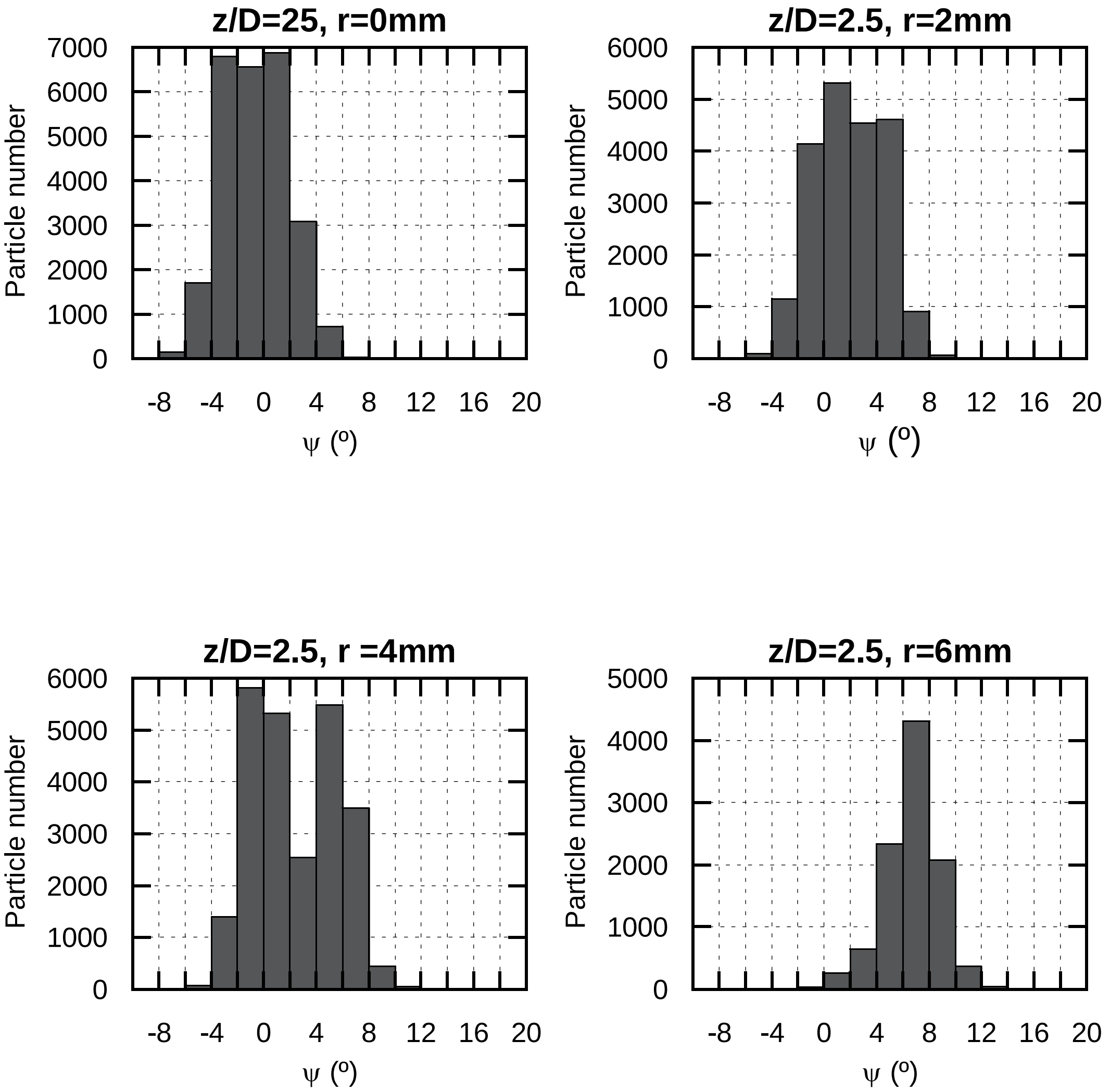

**Fig 18 a, b, c, d. Directional distribution for some points at the z=2.5D section**

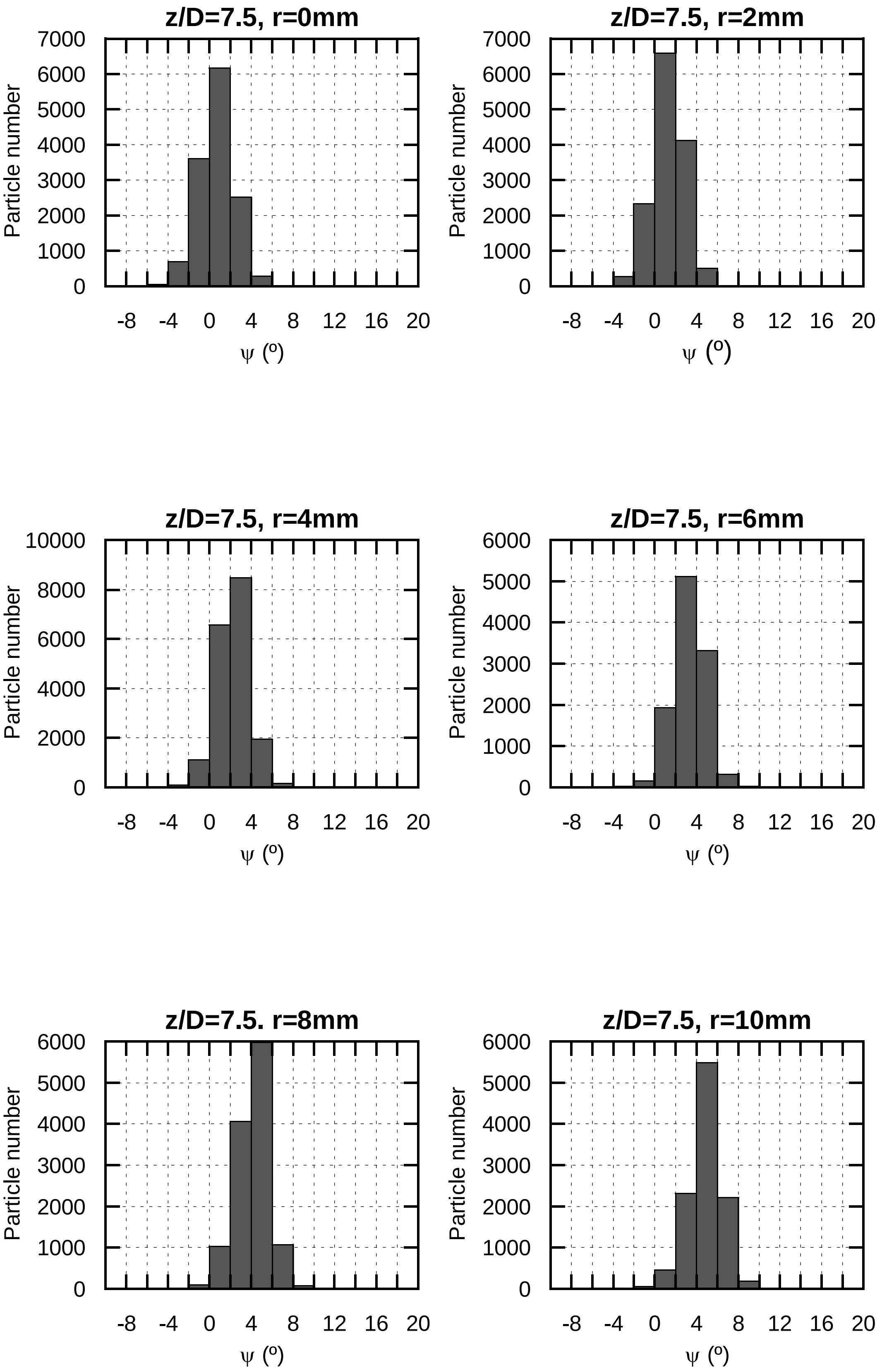

**Fig 19 a, b, c, d, e, f. Directional distribution for some points at the z=7.5D section**

All the distributions in the z=7.5D section show a unique peak, but for the z=2.5D section there are some points with a distribution with two local maxima. This appears associated to the focalization of particles. This focalization can start inside the nozzle for the greatest particles, but it continue even the z=2.5D section for the smallest particles. Directional classification removes the double-peaked shape and reduces the angle variance.

The drag force depends on the particle diameter. To see the variation of the velocity profiles with the diameter, we have split the size distribution in five intervals. We have used the classes: 50-60 μm, 60-70 μm, 70-80 μm, 80-90 μm and 90-100 μm. For simplicity, only data of the smallest (50-60 μm), the middle (70-80 μm) and the biggest (90-100 μm) are provided. Figures 20a, b, c, d show the mean axial velocity profile for these classes.

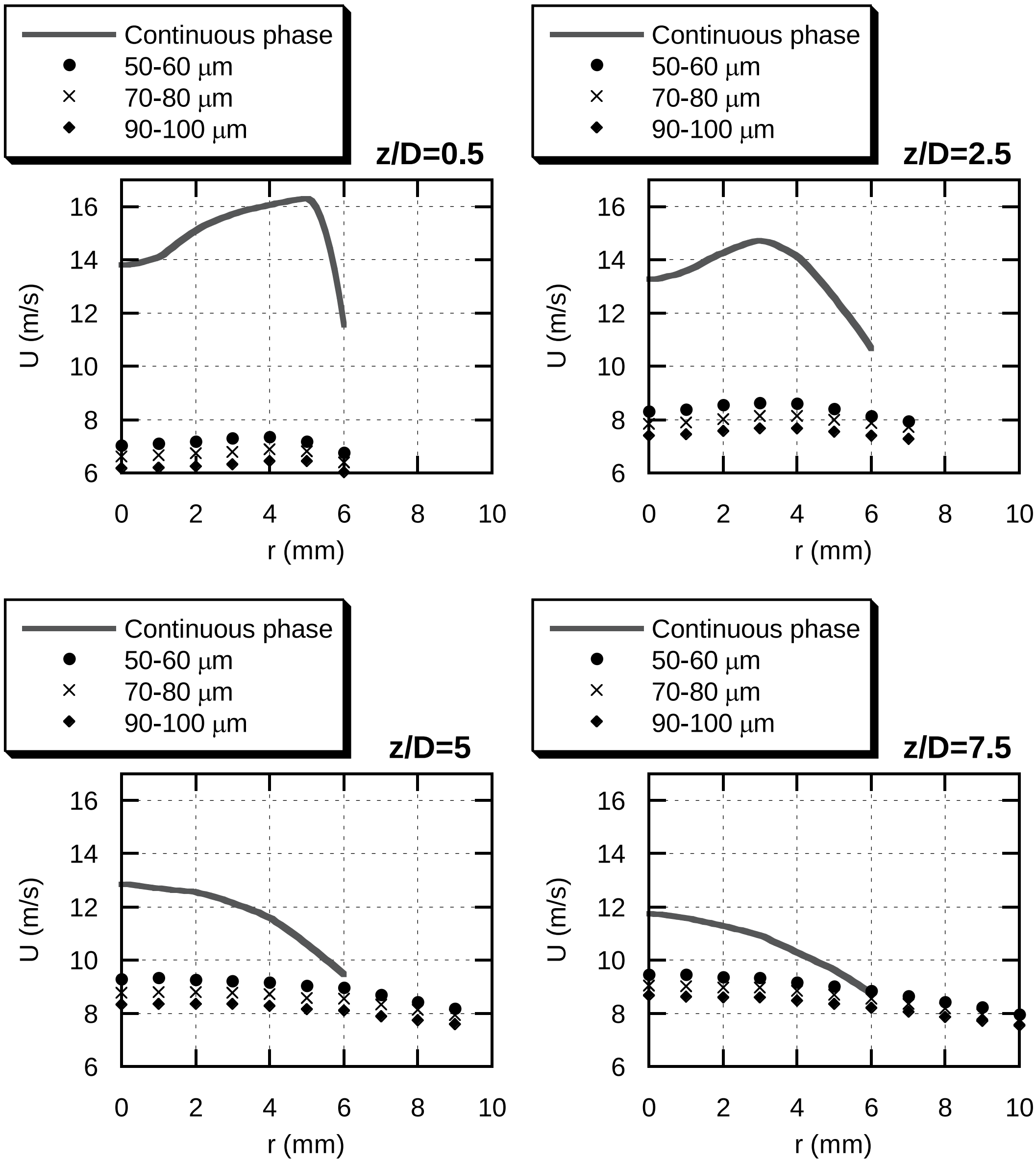

**Fig 20 a, b, c, d. Continuous phase axial mean velocity profiles for the 50-60 μm, 70-80 μm and 90-100 μm classes.**

Following Klyachko, the variation of particle velocity due to aerodynamic drag can be evaluated as:

$$\frac{dU_p}{dt} = \frac{18\mu}{D_p^2 \rho_p} \left[ 1 + \frac{1}{6} \left( \frac{\rho_g D_p |U_r|}{\mu} \right)^{2/3} \right] U_r$$

Where: $\mu$ is the dynamic viscosity; $D_p$ is the particle diameter; $U_r$ the relative velocity between gas and particle; $\rho_g$ , $\rho_p$ the gas and particle densities.

As the theory predicts, the smaller particles have the larger velocity because they follow better the continuous phase, with a larger velocity. However, the shape of all profiles is convex for the same section ( $z = 5D$ ) for all the diameter classes. In the central zone the gas increases de particle velocity for all section, but in the external layers the gas produces a decrease of the particle velocity in the last sections.

The comparison with the continuous phase shows that at the section z=5 and 7.5 D the particles in the central zone are accelerated by the gas, but air decelerates the particles in the external layers.

Finally, we have obtained the volume concentration for each diameter class. This is the information that appears in the figures 21a, b, c, d.

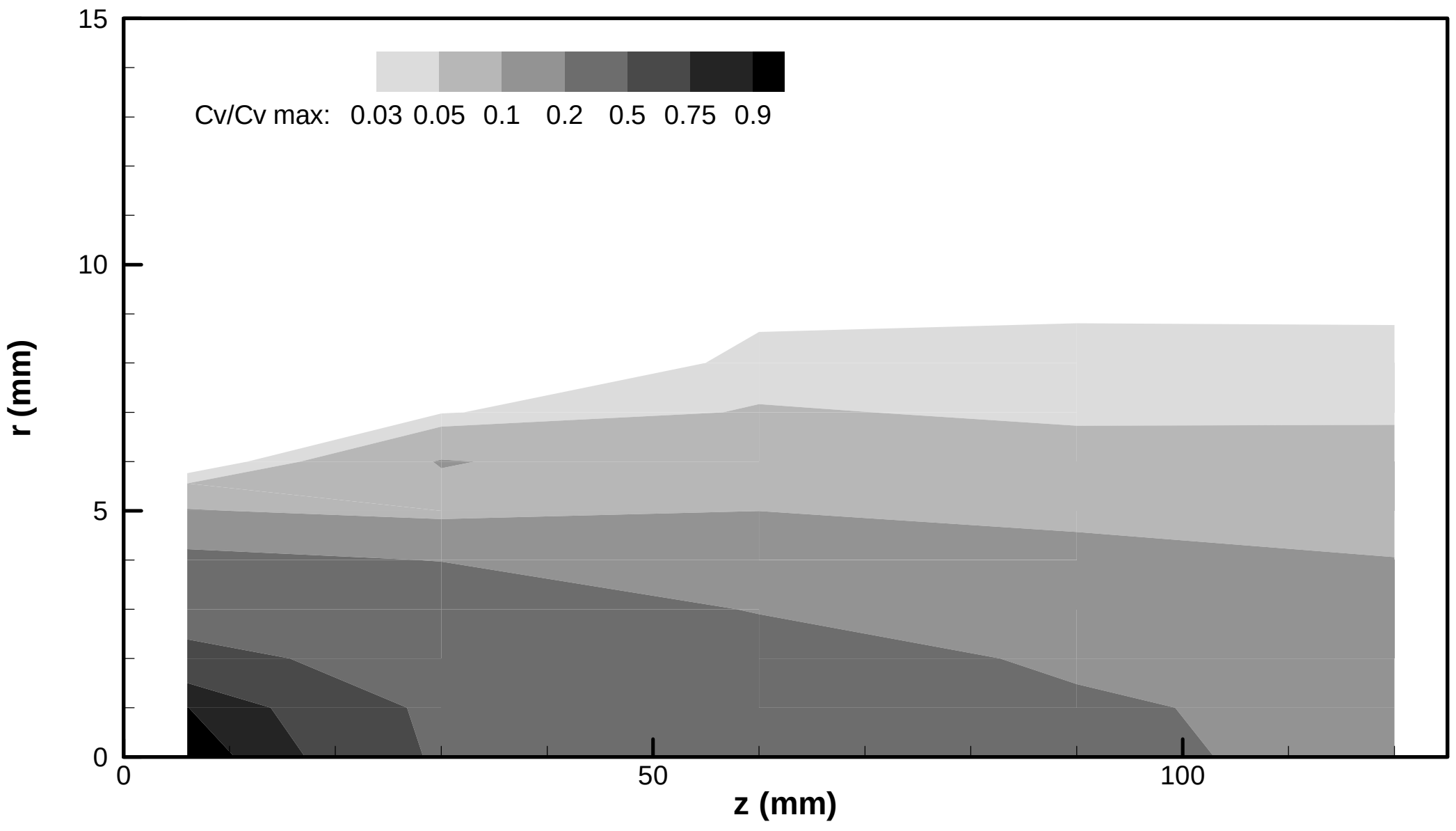

**Fig 21 a. Concentration map for the 50-60 μm.**

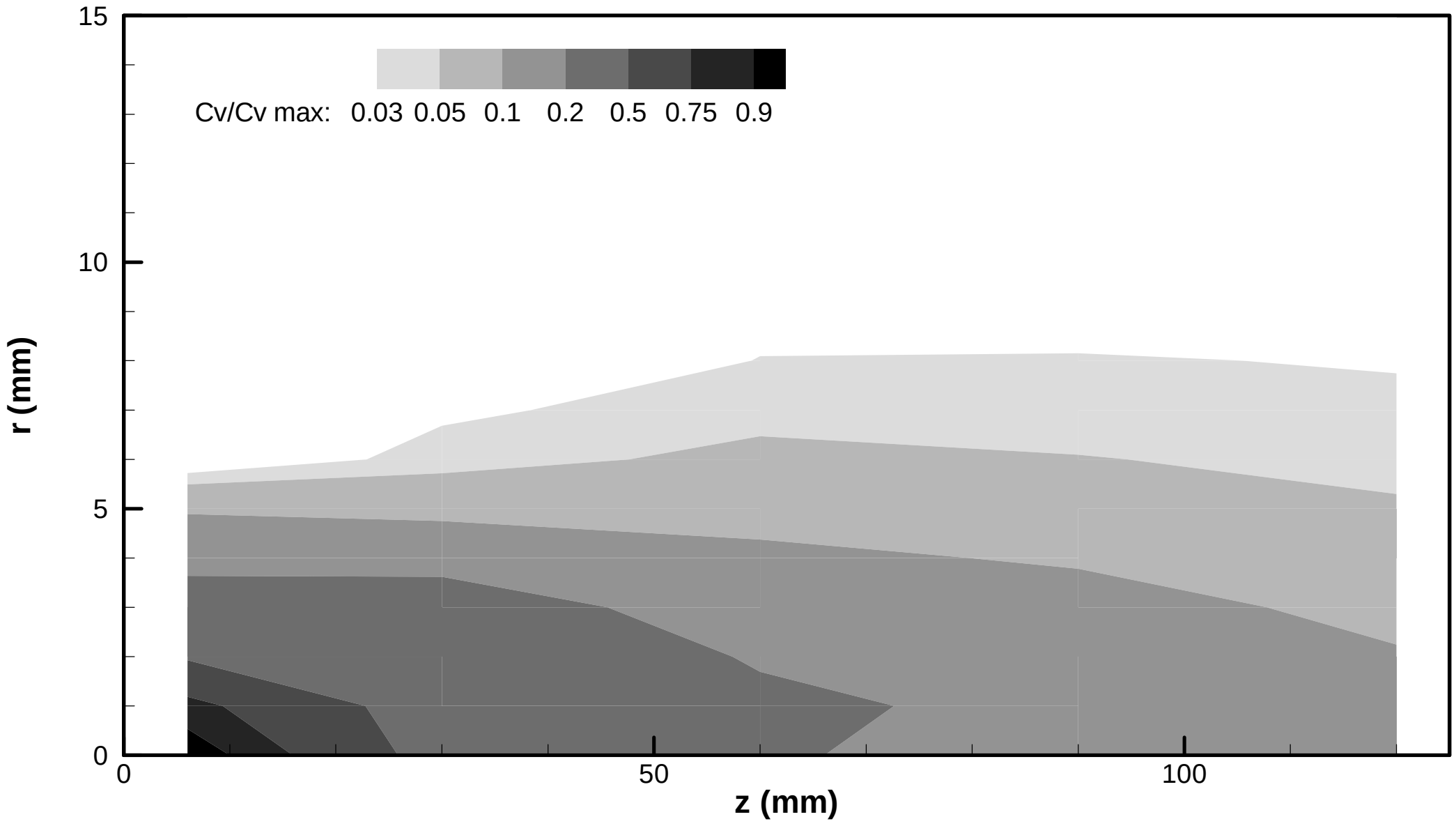

**Fig 21 b. Concentration map for the 70-80 μm.**

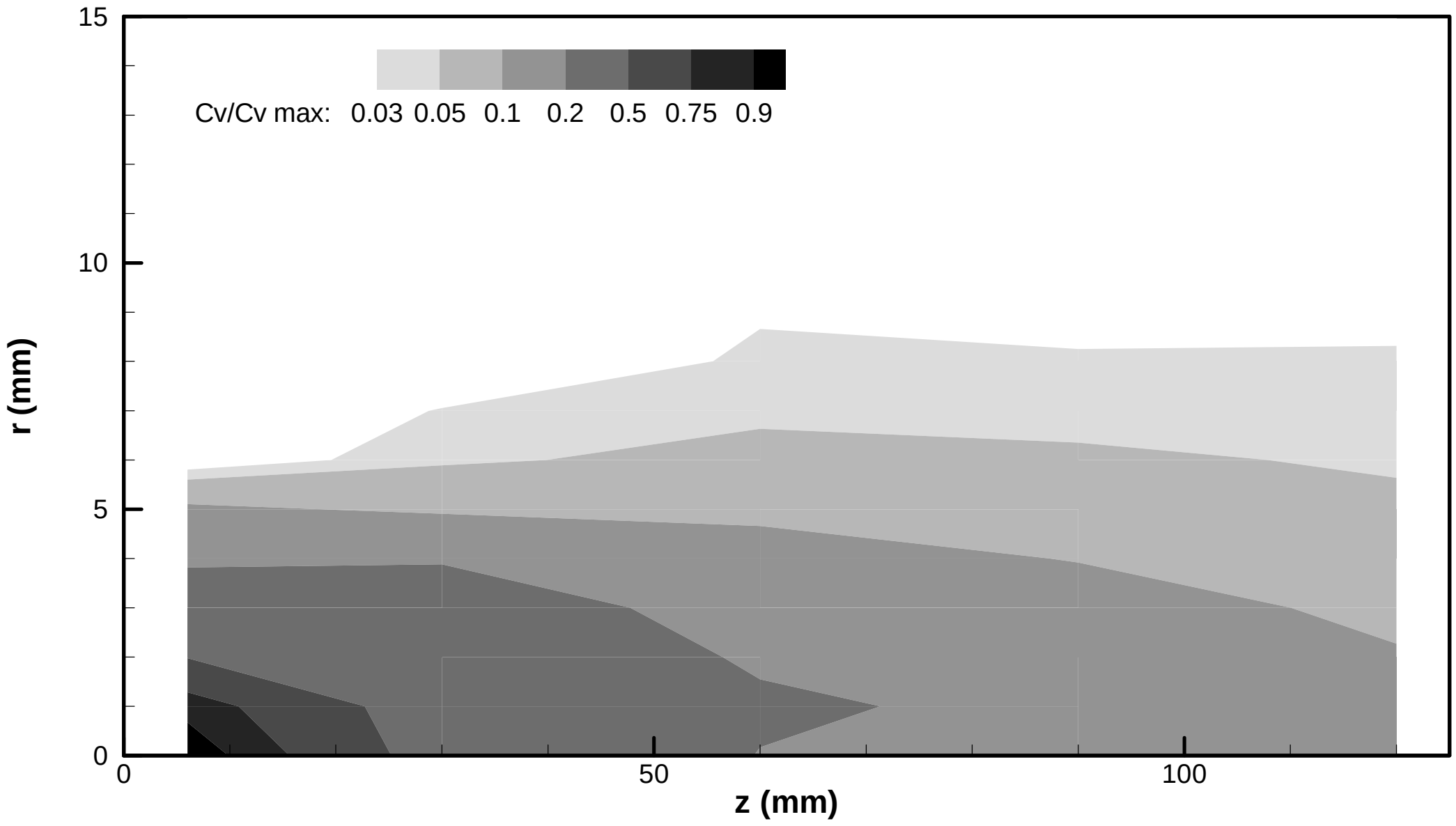

**Fig 21 c. Concentration map for the 90-100 μm.**

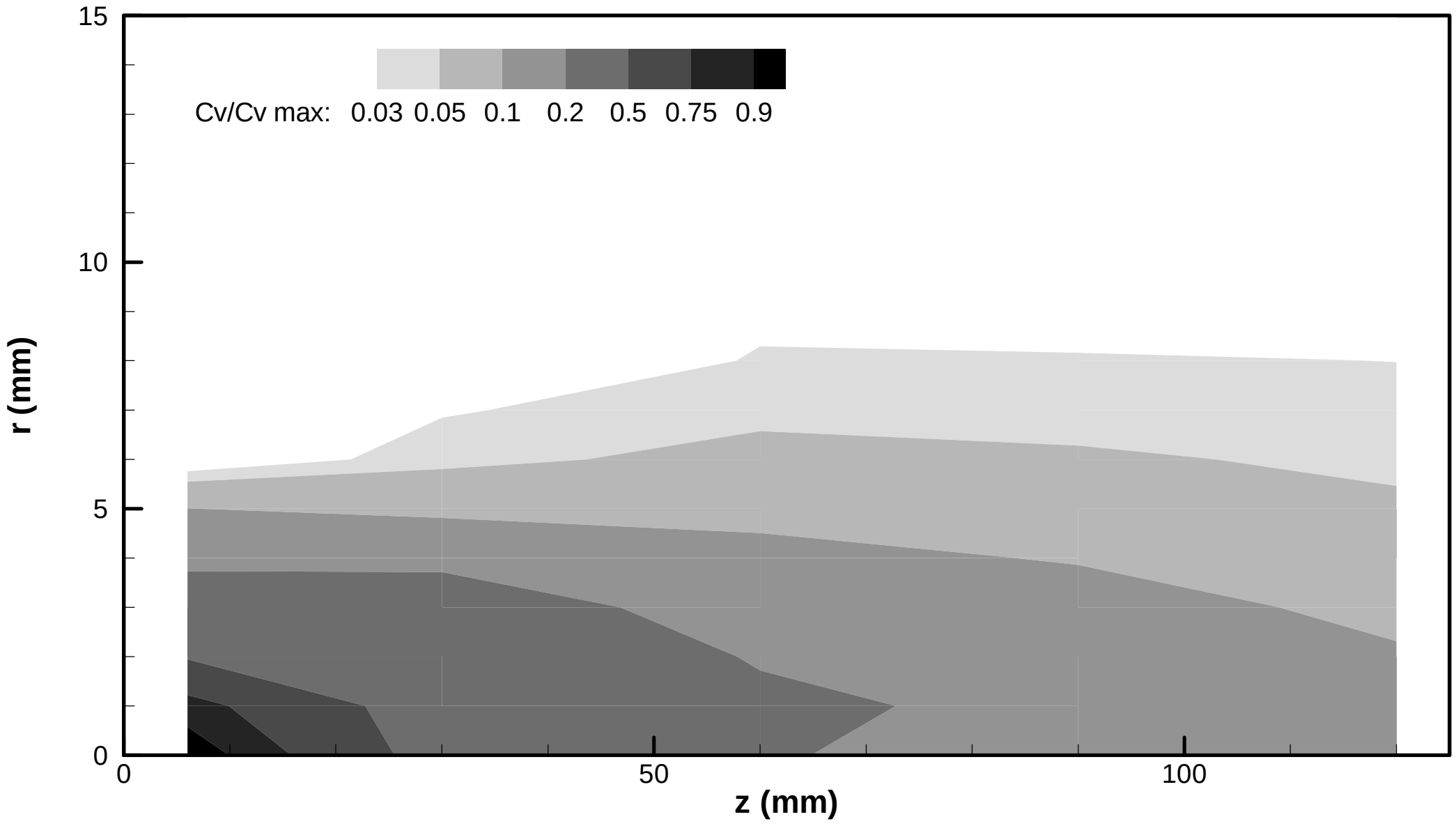

**Fig 21 d. Concentration map for the whole distribution.**

The evolution of the concentration in the axis can be seen in the figure 22 in a dimensionless way (the local concentration divided by the concentration in z=0.5D):

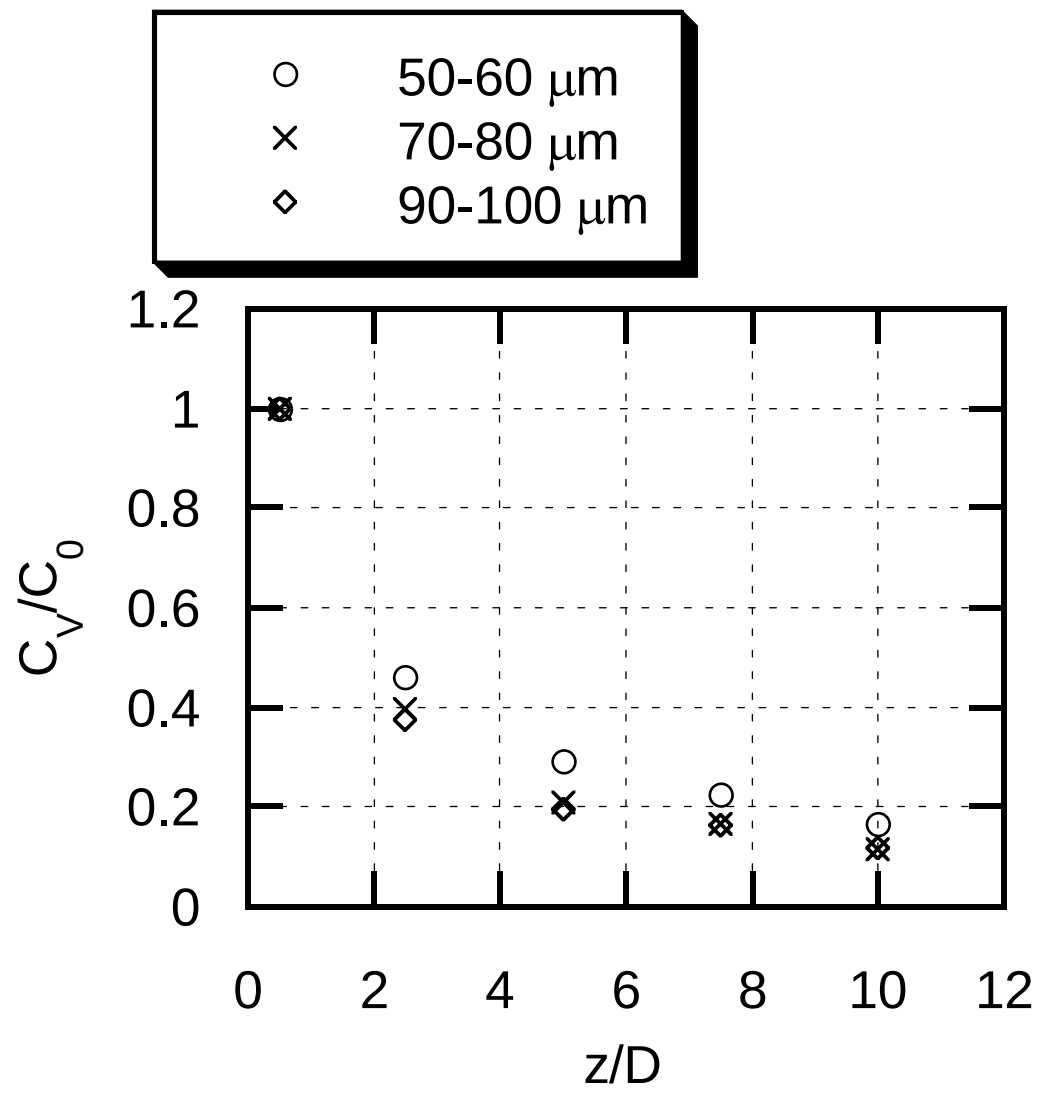

**Fig 22. Axial evolution of dimensionless concentration.**

The behavior is very similar for all sizes classes but the concentration decaying for the smallest diameter classes is slightly slower. This result suggests that the smallest particles focus downstream than the biggest ones.

*Far to the nozzle zone:*

*A) Continuous phase.*

The axial mean velocity of the continuous phase is shown in the sections $z = 10D, 20D, 30D \text{ and } 40D$ in the figure 23a.

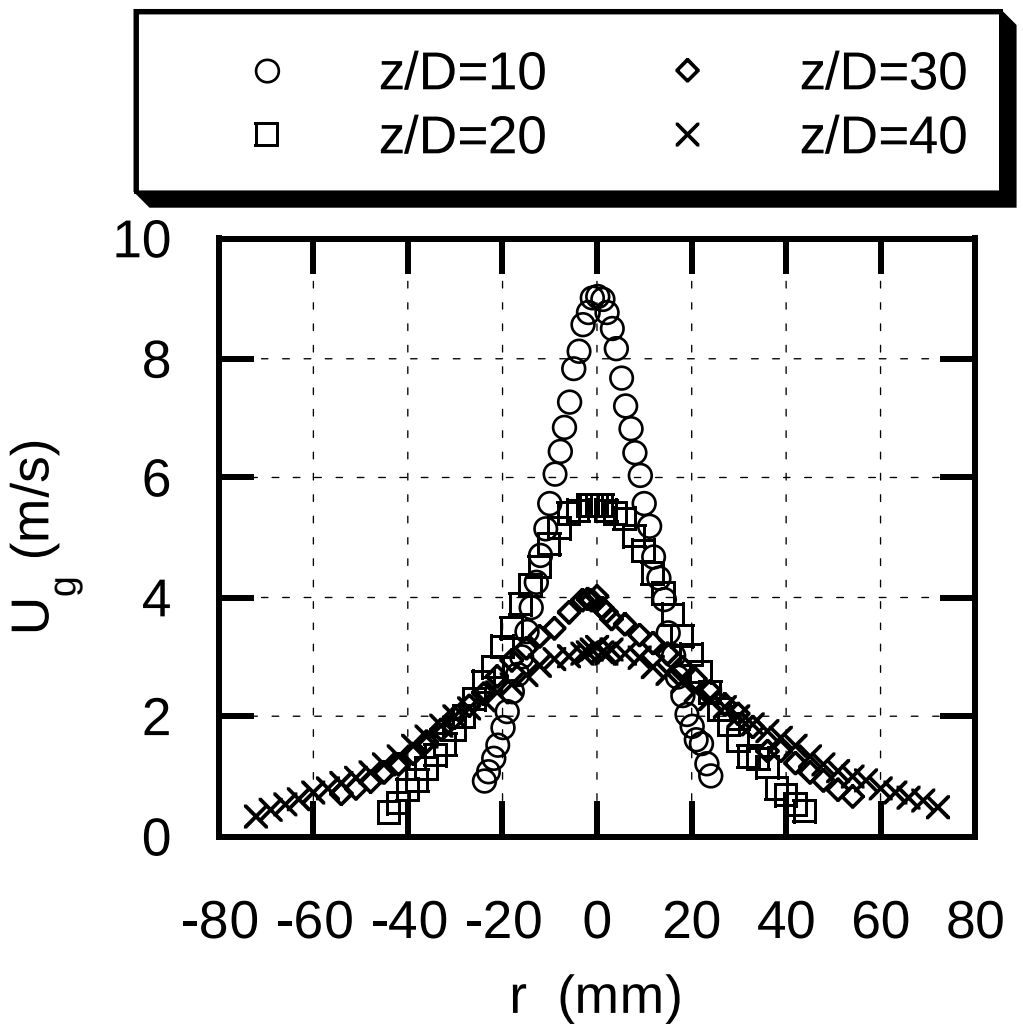

**Fig 23 a. Mean axial velocity for the continuous phase.**

In the figure, it can be noted the self similarity of the velocity profiles. This phenomenon is given in the single phase jet also starting from the section $z = 20D$.

We can characterize the expansion jet using the distance $b_{1/2}$. This distance is defined as the radial position at which the velocity is a half of the axis velocity.

The same parameter has been calculated for the single phase jet and for the particles. The next graph shows the results:

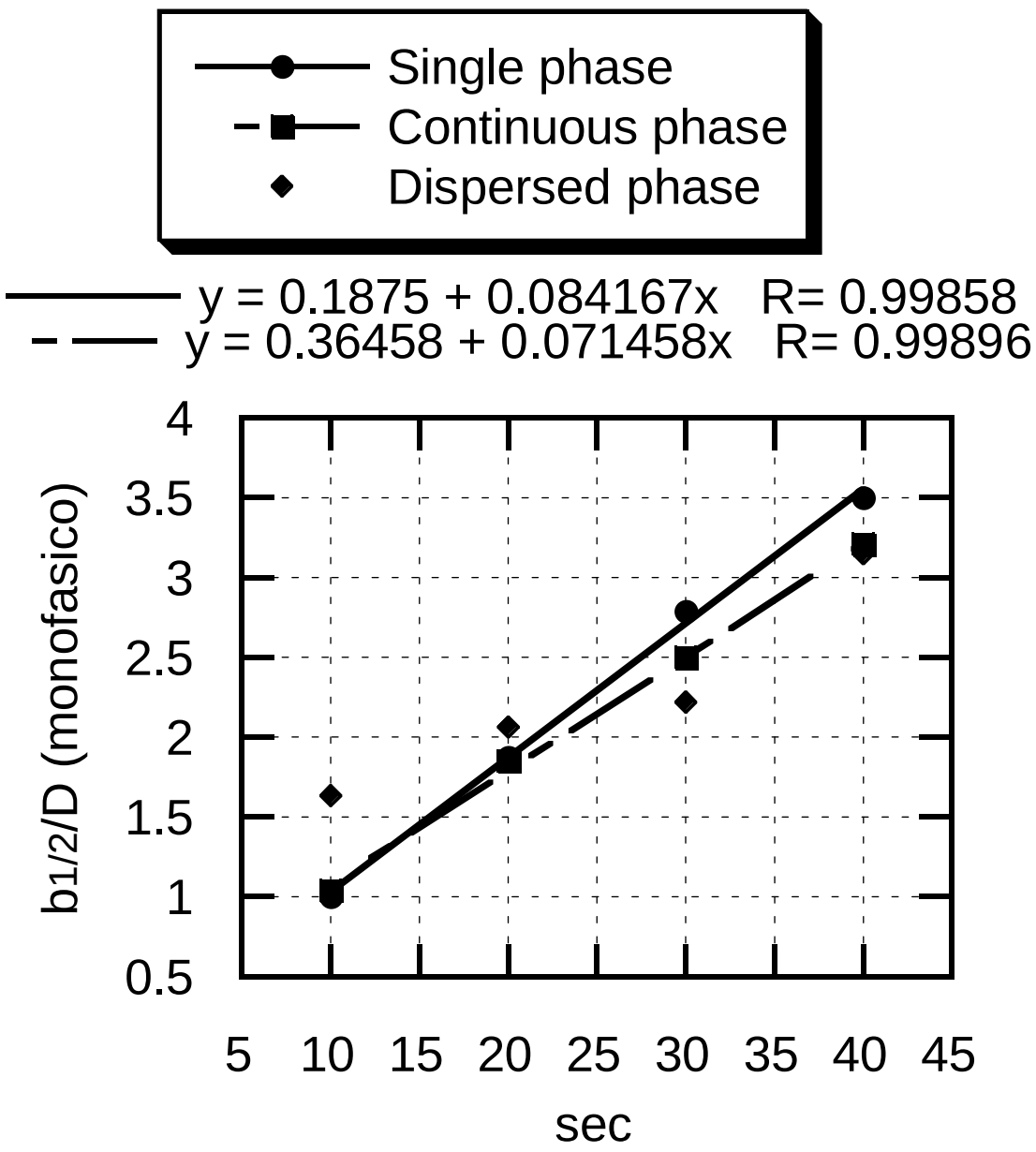

**Fig 24. Evolution of the $b_{1/2}$ parameter.**

The dispersed phase $b_{1/2}$ does not show a linear evolution. However, it is possible the comparison between the single phase jet and the continuous phase. The particle load induce a slower expansion rate starting from z=10D (in this axial position the parameter $b_{1/2}$ is similar).

*B) Dispersed phase.*

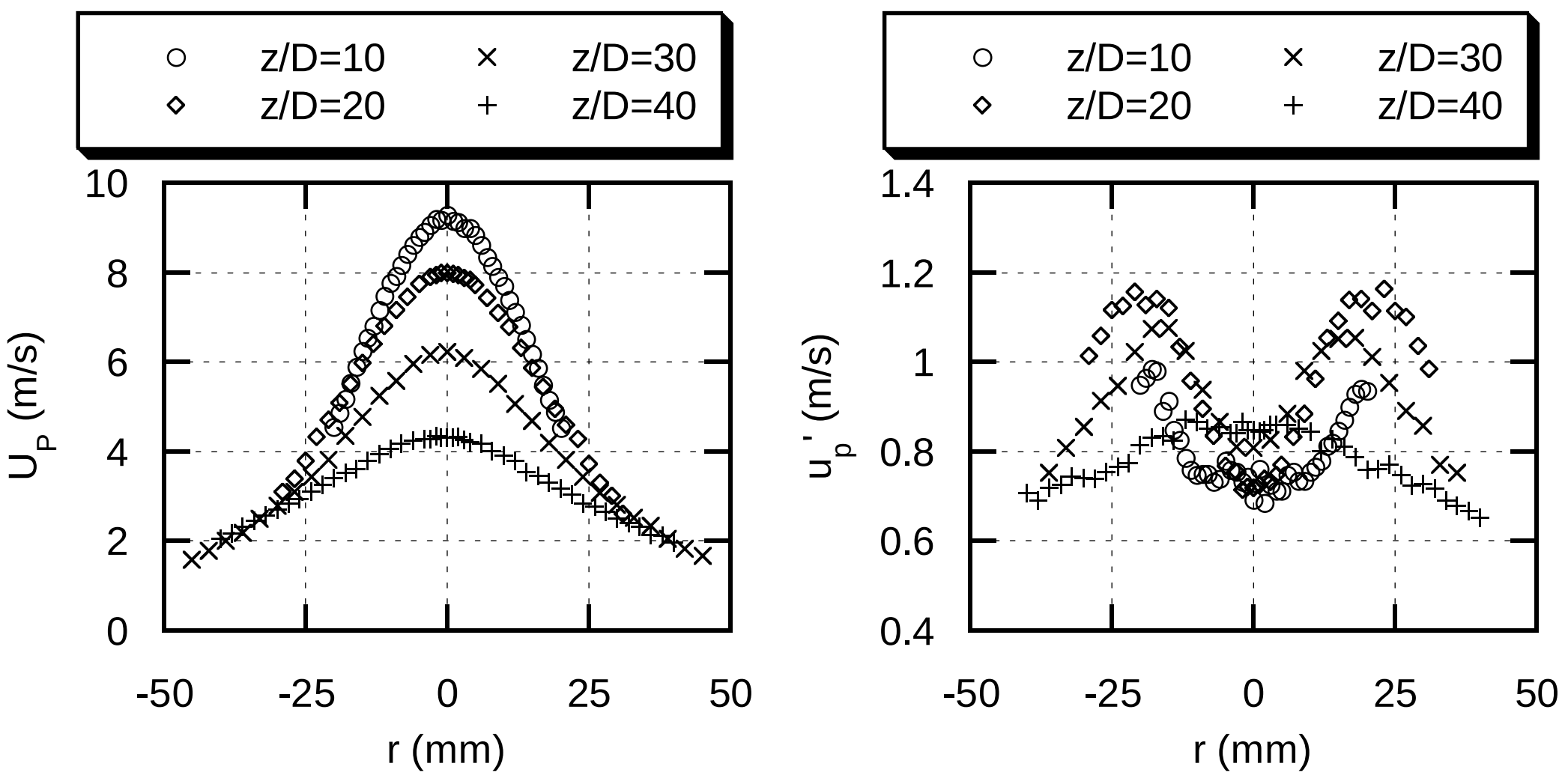

**Fig 25 a, b. Particle mean velocity and its fluctuation.**

The $b_{1/2}$ parameter shows a change of behavior between the first sections of the advanced zone and the z=40D section. Also, the axial fluctuation profile changes its shape for the z=40D section. It suggests a change between a development driven by the mean flow for the first sections and a regime affected by the turbulent at the last studied zone.

For the section $z = 20D$, it is possible to find data from other authors. The next graph shows the comparison with experimental results of Prevost (1994) and Modarres, Tan and Elghobashi (1983). For this section, we have made measurements with a mass ratio of 0.9 too.

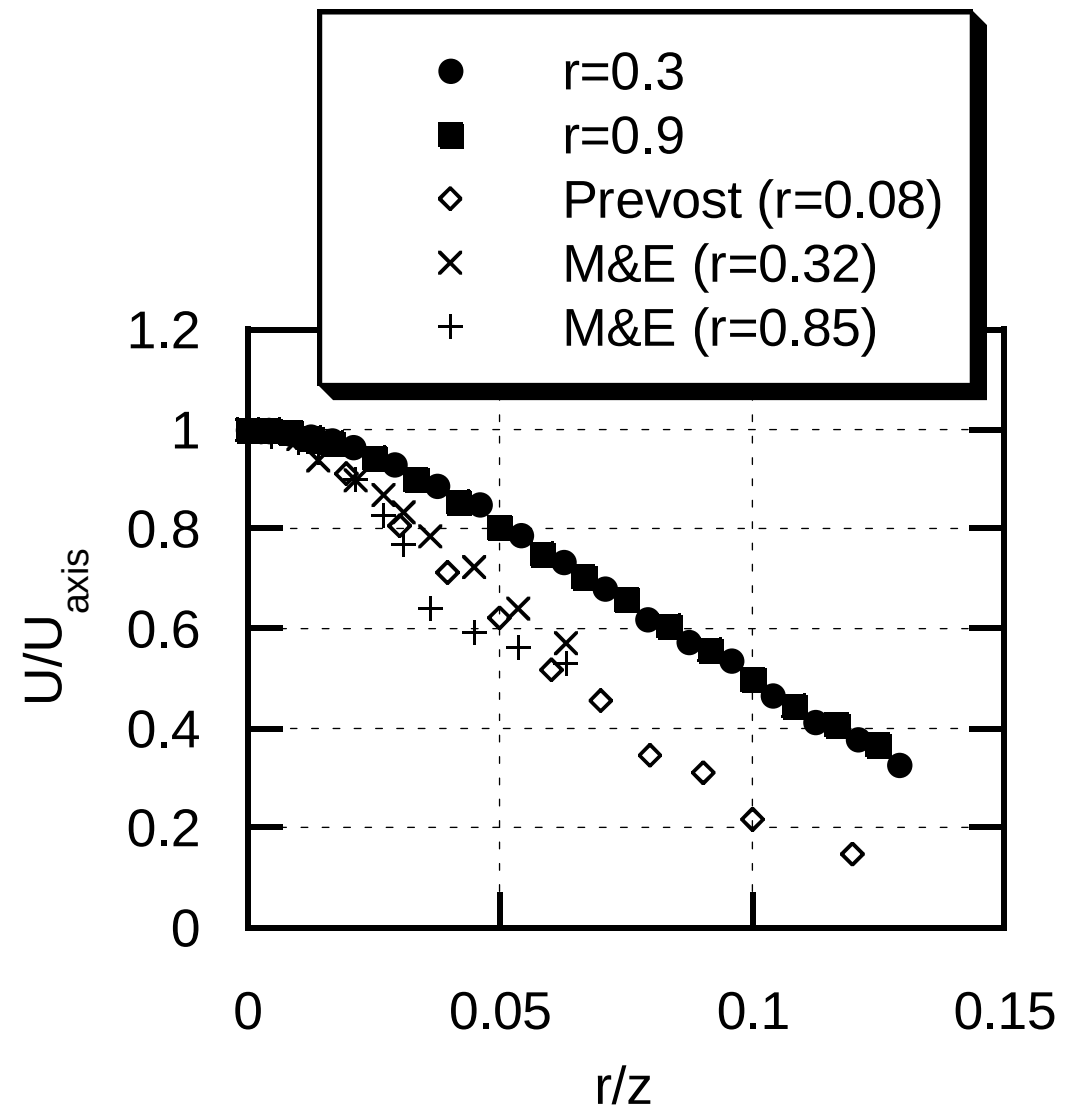

**Fig 26. Comparison at z=20D of the mean axial velocity of particles.**

The graph shows that velocity profile shape is similar for the cases whose injection is carried out with a pipe. However, with a convergent nozzle injection, the velocity profile is wider, and similar for the two tested mass ratios.

## *CONCLUSIONS*

A particle laden jet has been experimentally described. As a main characteristic, this jet is generated by a convergent nozzle. Special care has been taken with the control of the initial conditions (gas mass flow and mass ratio between particles and gas). The measured region includes zones of Stokes number medium-high.

The experimental characterization includes velocity measurements (axial evolution and transversal profiles) and the close to nozzle zone description is completed with mass flux and concentration profiles.

The study shows as the nozzle geometry change the initial conditions in comparison with other works carried out with pipe. The collisions of the particles with the nozzle wall induce a focalization effect that determines the evolution in the near to the nozzle zone. Also, this modification in the initial conditions gives important changes in the behavior of the particles in advances zones of development, giving more important particle dispersion.

Other effects are showed as the modification of the continuous phase due to the particle presence or the particle behavior modification when the $S_t$ is close to 1.